\documentclass[useAMS,usenatbib]{mn2e}
\usepackage{times,graphicx,colordvi}
\newcommand{\apj}{ApJ}                 
\newcommand{\mnras}{MNRAS}             \newcommand{\aap}{A\&A}
             
\newcommand{\aj}{AJ}                   \newcommand{\na}{New Astronomy}
              
    \newcommand{\pre}{Phys Rev E}
               \newcommand{\physrep}{Physics Reports}
\newcommand{\dd}{{\rm d}}
\newcommand{\vc}[1]{\textbf{\emph #1}}
\begin{document}
\label{firstpage}

\title[Statistical mechanics for self-gravitating systems]
{Equilibrium statistical mechanics for self-gravitating systems: local ergodicity and extended Boltzmann-Gibbs/White-Narayan statistics}
\author[P. He]{Ping He\thanks{Email: hep@itp.ac.cn} \\
Key Laboratory of Frontiers in Theoretical Physics, Institute of Theoretical Physics, Chinese Academy of Sciences, Beijing 100190, China}
\date{\today}
\maketitle

%-------------------------------------------------------------------------------------
\begin{abstract}
The long-standing puzzle surrounding the statistical mechanics of self-gravitating systems has not yet been solved successfully. We formulate a systematic theoretical framework of entropy-based statistical mechanics for spherically symmetric collisionless self-gravitating systems. We use an approach that is very different from that of the conventional statistical mechanics of short-range interaction systems. We demonstrate that the equilibrium states of self-gravitating systems consist of both mechanical and statistical equilibria, with the former characterized by a series of velocity-moment equations and the latter by statistical equilibrium equations, which should be derived from the entropy principle. The velocity-moment equations of all orders are derived from the steady-state collisionless Boltzmann equation. We point out that the ergodicity is invalid for the whole self-gravitating systems, but it can be re-established locally. Based on the local ergodicity, using Fermi-Dirac-like statistics, with the nondegenerate condition and the spatial independence of the local microstates, we rederive the Boltzmann-Gibbs entropy. This is consistent with the validity of the collisionless Boltzmann equation, and should be the correct entropy form for collisionless self-gravitating systems. Apart from the usual constraints of mass and energy conservation, we demonstrate that the series of moment or virialization equations must be included as additional constraints on the entropy functional when performing the variational calculus; this is an extension to the original prescription by White \& Narayan. Any possible velocity distribution can be produced by the statistical-mechanical approach that we have developed with the extended Boltzmann-Gibbs/White-Narayan statistics. Finally, we discuss the questions of negative specific heat and ensemble inequivalence for self-gravitating systems.
\end{abstract}

%--------------------------------------------------------------------------------------
\section{Introduction}
\label{sec:intro}

It is known that conventional methods for the statistical mechanics of short-range interaction systems cannot be used directly to study long-range self-gravitating systems. Hence, it is necessary to return to the starting point of statistical mechanics and to develop special techniques to handle the long-range nature of gravity \citep{padmanabhan90}. However, for more than several decades, the statistical mechanics of collisionless self-gravitating systems has not been realized successfully, and it still remains a puzzle \citep*[see, for example,][]{ogorodnikov57, ant62, lb67, shu78, tremaine86, sridhar87, stiavelli87, white87, spergel92, soker96, kull97, chavanis98, nakamura00, hansen05, levin08a, hjorth10}.

In a previous work \citep{hep10}, we performed a preliminary investigation of the statistical mechanics for self-gravitating systems. We employed a phenomenological entropy form of ideal gas, first proposed by \citet{white87}, to revisit this question. By calculating the first-order variation of the entropy, subject to the usual mass- and energy-conservation constraints, we obtained an entropy stationary equation. Combined with the Jeans equation, and by specifying some functional form for the anisotropy parameter $\beta$, we numerically solved the two equations. We demonstrated that the velocity anisotropy parameter plays an important role in attaining a density profile that is finite in mass, energy and spatial extent. If combined again with some empirical density profile from simulations, our theoretical predictions of the anisotropy parameter, and the radial pseudo-phase-space density $\rho/\sigma^3_r$ in the outer regions of the dark matter halo, agree very well with the data of $N$-body simulations. The predictions are also acceptable in the middle regions of dark halos. Disagreements occur in the inner regions, probably because some important physics has been ignored.

The second-order variational calculus reveals the seemingly paradoxical, but actually complementary, consequence that the equilibrium state of self-gravitating systems is not only the global minimum entropy state for the whole system -- under the condition of energy conservation and the assumption that $\rho$ is an unknown variable function, whereas the two second-order velocity moments $\rho\overline{v^2_r}$ and $\rho\overline{v^2_t}$ are known invariant functions of the radius $r$ -- but also, simultaneously, the local maximum entropy state for every and any small part of the system. This finding suggests that equilibrated self-gravitating systems are not in the maximum entropy states, and the local statistical properties of self-gravitating systems should be different from their global properties.

In a follow-up work \citep{kang11}, by introducing an effective pressure instead of a radial pressure to construct the specific entropy, we used the entropy principle. We proceeded in a similar way as in \citet{hep10} and obtained a new entropy stationary equation. An equation of state for equilibrated dark haloes is derived from this entropy stationary equation, from which the dark halo density profiles can be obtained. We also derived the anisotropy parameter and pseudo-phase-space density profile. All these predictions agree well with numerical simulations in the outer regions of dark haloes. This work provides further support to the idea that statistical mechanics of self-gravitating systems is feasible. We believe that our findings might provide crucial clues for the development of statistical mechanics for both self-gravitating systems and other long-range interaction systems.

However, our treatments suffer from the following drawbacks. (i) The entropy forms are only chosen phenomenologically, without robust physical origins. (2) We calculate the first-order variation only with respect to the density profile, but we treat other variables, such as the pressure, as known functions. Thus, the variational calculus is incomplete. (3) We could still miss some crucial physics. Nevertheless, from these works, we can still draw the following conclusions.
% --------------------------------------------------------------------------------------
\begin{enumerate}
\item Both the concept of entropy and entropy principle are still valid for self-gravitating systems. This finding implies that the entropy-based statistical mechanics of self-gravitating systems is also valid, and we do not agree that a dynamical theory can account for the coarse-grained phase-space density \citep[cf.][]{lb05}.

\item Entropy is extensive. We first constructed the specific entropy, and then obtained the total entropy in the additive and hence extensive way, as in equation~(4) of \citet{hep10} or equation~(6) of \citet{kang11}. The fact that such an extensively defined entropy works very well indicates that a non-extensive entropy, such as the entropy of \citet{tsallis88}, is not necessary.

\item The equilibrium states also consist of mechanical equilibrium. In \citet{hep10} and \citet{kang11}, we combined the entropy stationary equation with the Jeans equation to predict the final state of the system. This proved to be a successful approach. The success of this treatment indicates that the final states are not only statistical equilibrium states, but also mechanical equilibrium states.

\item Local and global equilibria are different. That is, for an equilibrated self-gravitating system, the local statistical properties differ from the global properties. This difference suggests that we should develop the statistical mechanics of self-gravitating systems in a different way from the standard statistical mechanics of short-range interactions.
\end{enumerate}

We use the above four principles to formulate the theoretical framework of the statistical mechanics for spherically symmetric collisionless self-gravitating systems. The paper is organized as follows. In Section~\ref{sec:relaxme}, we present the series of moment equations, which are  derived from the spherical collisionless Boltzmann equation (CBE) and are employed to characterize the mechanical equilibria of the system. In Section~\ref{sec:ergodic}, we demonstrate that ergodicity is invalid globally for self-gravitating systems, but that it can be re-established locally if the particles are treated as indistinguishable. In Section~\ref{sec:core}, we present the entire framework of the statistical mechanics, with a truncated distribution function (DF) to show how it is possible to work with this statistical mechanics. We give a discussion and conclusions in Section~\ref{sec:concl}.

%------------------------------------------------------------------------------------
\section{Relaxation processes and mechanical/virial equilibria}
\label{sec:relaxme}
%---------------------------------------------------------------------------------------
\subsection{Relaxation Processes}
\label{sec:relax}

In physical dynamics, relaxation refers to the essential process by which a system approaches equilibrium or by which a perturbed system returns to equilibrium \citep*{mo10}. So far, four collisionless relaxation processes have been identified as being responsible for the equilibrium state of self-gravitating systems: (i) violent relaxation, (ii) Landau damping, (iii) phase mixing and (iv) chaotic mixing. We refer the reader to \citet{mo10} for a detailed discussion of these relaxation processes. There could also be other possible relaxation mechanisms \citep[e.g.][]{gurzadyan86}.

% Moment equations and mechanical equilibriums
%------------------------------------------------------------------------------------
\subsection{Velocity moment equations and mechanical equilibria}
\label{sec:me2}

The kinetic-theoretical description of self-gravitating systems is well approximated by the CBE as
% -------------------------------------------------------------------------------------
\begin{equation}
\label{eq:cbe}
\frac{\dd f}{\dd t} = \frac{\partial f}{\partial t} + \vc{v} \cdot \frac{\partial f}{\partial \vc{x}} - \frac{\partial \Phi}{\partial \vc{x}} \cdot \frac{\partial f}{\partial \vc{v}} = 0.
\end{equation}
Here, $f(t, \vc{x}, \vc{v})$ is the fine-grained statistical DF. The CBE is an approximation that results from the Liouville equation when the particle number of the system is very large, and the $N$-body DF can be separated as the product of single-particle DFs as
% --------------------------------------------------------------------------------------
% N-body distribution function
\begin{equation}
\label{eq:nbdf}
f^{(N)}(t,\vc{w}_1, \ldots, \vc{w}_N)=\prod^N_{i=1} f(t,\vc{w}_i).
\end{equation}
Here, $\vc{w}_i = (\vc{x}_i, \vc{v}_i)$ denotes the phase-space coordinates for the $i$-th particle. This assumption implies that positions of gravitating particles are uncorrelated \citep[i.e. the probability of finding a particle near any phase-space position is unaffected by the presence or absence of particles at nearby points; see][]{galdyn08}. We refer to this separability of the $N$-body DF as the statistical independence of particle positions. Furthermore, the gravitational effect exerted on a specific particle is collectively represented by the mean gravitational potential $\Phi(\vc{x})$, contributed from all the particles of the system. This mean background potential $\Phi(\vc{x})$, as well as the uncorrelation assumption, suggests that the direct gravitational interaction between any two particles is unimportant, and hence should be neglected (i.e. the particles among themselves are collisionless).

The CBE might not be a good approximation in very dense regions, such as the center of a dark matter halo, where two-body relaxation might play an important role. In this case, the CBE would be better replaced by, say, the Balescu-Lenard equation\footnote{However, the two-body relaxation was overemphasized in \citet{hep10}.} \citep{heyvaerts10, chavanis10}.

Throughout this work, spherical symmetry is always assumed for self-gravitating systems. However, we believe that it is not difficult to extend our results to general cases. We multiply the steady-state CBE with various powers of velocity, and we integrate over all velocities to obtain the moment equations of all orders (see Appendix~\ref{sec:appdx}). The second-order equation is the familiar Jeans equation:
% --------------------------------------------------------------------------------
\begin{equation}
\label{eq:2ndoe}
\frac{\dd}{\dd r}(\rho\overline{v^2_r}) + \frac{2\rho}{r}( \overline{v^2_r} - \overline{v^2_t}) = \frac{\dd}{\dd r}(\rho\overline{v^2_r}) + \frac{2\beta}{r} \rho\overline{v^2_r} = - \rho \frac{\dd \Phi}{\dd r}.
\end{equation}
Here, $v_t$ denotes either $v_{\theta}$ or $v_{\phi}$ and the barred quantities indicate the corresponding velocity moments. $\beta$ is the usual velocity anisotropy parameter, defined as $\beta = 1 - \overline{v^2_t}/\overline{v^2_r}$. We can also obtain two 4th-order equations as
% -------------------------------------------------------------------------------------
\begin{equation}
\label{eq:4thoe1}
\frac{\dd}{\dd r}(\rho\overline{v^4_r}) + \frac{2\rho}{r}(\overline{v^4_r} - 3 \overline{v^2_r v^2_{t}}) = - 3\rho\overline{v^2_r}\frac{\dd \Phi}{\dd r},
\end{equation}
and
\begin{equation}
\label{eq:4thoe2}
\frac{\dd}{\dd r}(\rho\overline{v^2_r v^2_{t}}) + \frac{4\rho}{3 r} (3\overline{v^2_r v^2_{t}} - \overline{v^4_{t}}) = - \rho\overline{v^2_{t}}\frac{\dd \Phi}{\dd r}.
\end{equation}
These two equations are actually the same as those of \citet{merrifield90}, except that \citet{merrifield90} used $v_t$ to denote the two-dimensional velocity component, as $v_t = (v^2_{\theta} + v^2_{\phi})^{1/2}$. The general moment equations are given by equations~(\ref{eq:meq2}) \citep[see also][]{dejonghe92, an11},
% -------------------------------------------------------------------------------------
\begin{displaymath}
\frac{\dd}{\dd r}(\rho\overline{v^{k+2}_r v^m_{t}}) + \frac{(m+2)}{r} \rho\overline{v^{k+2}_r v^m_{t}} - \frac{(k+1)}{r}\frac{(m+2)}{(m+1)}
\end{displaymath}
\begin{displaymath}
{\hskip 5mm} \times \rho\overline{v^k_r v^{m+2}_{t}} = - (k+1)\frac{\dd \Phi}{\dd r} \rho\overline{v^k_r v^m_{t}}. {\hskip 23.25mm} \rm{(\ref{eq:meq2})}
\end{displaymath}
Generally, there are $N$ independent $2N$th-order moment equations, with different combinations of $k$ and $m$, in equation~(\ref{eq:meq2}). These moment equations characterize the complex mechanical equilibria of self-gravitating systems.

As demonstrated in Appendix~\ref{sec:appdx}, the DF $f$ is inseparable, and hence these infinitely many moments are independent and irreducible. The right-hand side of equation~(\ref{eq:meq2}), $-(k+1)\rho\overline{v^k_r v^m_{t}}\dd \Phi/\dd r$, indicates the self-coupling between the system and its own gravitational potential through various orders of velocity moments specified by different $k$ and $m$.

\citet{magorrian94} have shown that when moments up to the 10th-order have been determined, accurate predictions of the line-of-sight velocity distribution can be made. This result suggests that a complete and accurate prediction of the 3-dimensional DF might also need to be calculated to this high order.

%-------------------------------------------------------------------------------------
\begin{figure}
\centerline{\includegraphics[width=1.0\columnwidth]{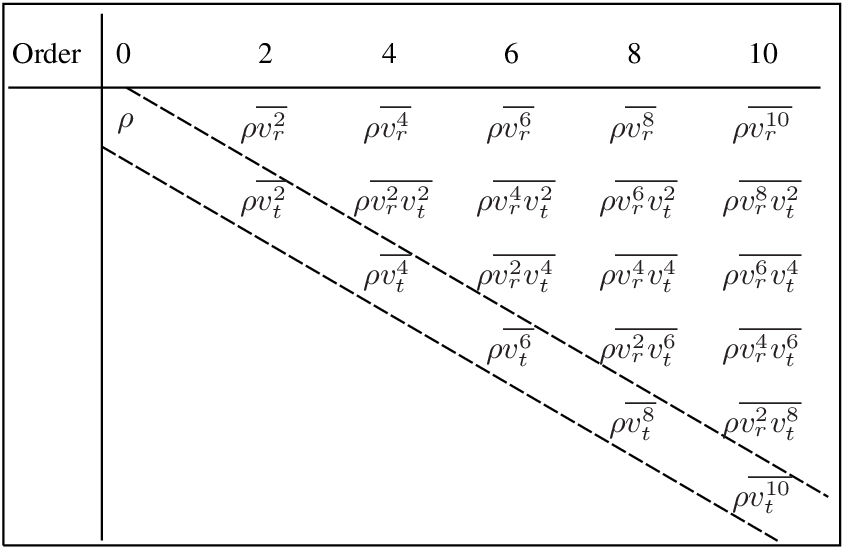}}
\caption{Independent velocity moments up to the 10th-order for spherically symmetric self-gravitating systems. The moments in the upper-right part are governed by the moment differential equations~(\ref{eq:meq2}), i.e. the moments appear in the differential terms. However, the moments $\rho\overline{v^k_t}$ in the diagonal, with $k=0, 2, 4, \ldots$, do not appear in the differential terms, and hence there are no differential equations for these.}
\label{fig:moments}
\end{figure}

% ------------------------------------------------------------------------------------
In Fig.~\ref{fig:moments} we show independent velocity moments up to the 10th-order for spherically symmetric self-gravitating systems. These non-vanishing moments are the effective thermodynamic variables for self-gravitating systems. Obviously, the number of equations of mechanical equilibria (i.e. the moment equations~\ref{eq:meq2}) is less than the number of independent moments. The series of moment equations~(\ref{eq:meq2}) does not constitute a closed equation system, because the moments $\rho\overline{v^k_t}$, the one-dimensional transverse moments of all orders ($k=0, 2, 4, \ldots$), do not appear in the differential terms, and hence there are no differential equations for them.

The DF $f$ that satisfies the CBE is the fine-grained DF, or phase-space density. Any physical quantity constructed with the fine-grained DF $f$, such as the entropy $S(f)$, should be conserved \citep[i.e. $\dd S / \dd t = 0$; see][]{mo10}. The meaningful DF for statistical mechanics is the so-called coarse-grained DF $f_c(\vc{ x},\vc{ v})$, usually defined as \citep[see][]{saslaw00, bindoni08, mo10}
% --------------------------------------------------------------------------------------
% coarse-grained DF-1
\begin{equation}
\label{eq:cgdf1}
f_c(\vc{x},\vc{v}) \equiv \frac{1}{\Delta^6\vc{w}} \int_{\Delta^6 \vc{w}} f(\vc{ x}',\vc{v}') \dd^3\vc{x}'\dd^3\vc{v}'.
\end{equation}
This is the average of $f$ within a sufficiently small but not infinitesimal phase-space element, i.e. the macrocell, $\Delta^6\vc{w} = \Delta^3 \vc{x} \Delta^3\vc{v}$, centered at the phase point $(\vc{x}, \vc{v})$. By definition, $f_c$ always satisfies $f_c \leq f$. Also, it is not difficult to prove that the statistical independence suggested by equation~(\ref{eq:nbdf}) is also held for the coarse-grained DF.

With the definition of the coarse-grained DF, we can find the relationship between the coarse-grained velocity moment $\overline{Q_c}$ and the fine-grained moment $\overline{Q_f}$ of any quantity $Q(\vc{v})$, a smooth function of $\vc{v}$. We have
% -------------------------------------------------------------------------------------
\begin{equation}
\label{eq:qdf1}
\frac{1}{\Delta^6\vc{w}} \int_{\Delta^6 \vc{w}} Q(\vc{v}') f(\vc{ x}',\vc{v}') \dd^3\vc{x}' \dd^3\vc{v}' \simeq Q(\vc{v}) f_c(\vc{x},\vc{v}),
\end{equation}
where $Q(\vc{v}')\simeq Q(\vc{v})$ within the small phase-space element $\Delta^6\vc{ w}$, and we use the definition of the coarse-grained DF of equation~(\ref{eq:cgdf1}). Integrating both sides of equation~(\ref{eq:qdf1}) over \vc{v}, we obtain
% --------------------------------------------------------------------------------------
\begin{displaymath}
\overline{Q_c}(\vc{x})\equiv\int Q(\vc{v}) f_c(\vc{x}, \vc{v}) \dd^3\vc{v} = \frac{1}{\Delta^3 \vc{x}} \int_{\Delta^3 \vc{x}} Q(\vc{v}') f(\vc{ x}',\vc{v}') \dd^3\vc{x}' \dd^3\vc{v}'
\end{displaymath}
\begin{equation}
\label{eq:qdf2}
= \frac{1}{\Delta^3 \vc{x}} \int_{\Delta^3 \vc{x}} \overline{Q_f}(\vc{ x}')\dd^3\vc{x}' \simeq \overline{Q_f}(\vc{x}) + \frac{1}{24} \sum^3_{i=1} \frac{\partial^2 (\overline{Q_f})}{\partial x^2_i}\Delta x^2_i,
\end{equation}
where the last approximation is obtained by Taylor expansion with respect to \vc{x} to the second order. Hence, we can see that, although the coarse-grained DF does not obey the CBE, the coarse-grained moments differ from the fine-grained moments by only up to the second-order small quantities determined by the size of the finite volume element. So, we can safely use these moment equations to characterize the mechanical equilibria of the system, regardless of whether these moments are derived from the fine-grained or coarse-grained DF.

%--------------------------------------------------------------------------------
% virial theorem figure
\begin{figure}
\centerline{\includegraphics[width=\columnwidth]{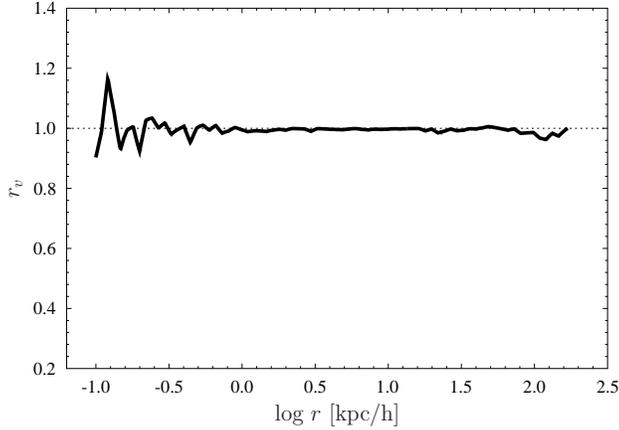}}
\caption{The virialization ratio $r_v$ vs. the radius $r$ of a virialized cold dark matter halo, which is calculated using equation~(\ref{eq:rv}). The simulation data are from \citet{navarro10} and \citet{ludlow10}. The result also implies that the CBE is a very good approximation for describing self-gravitating systems, such as dark matter halos.}
\label{fig:vratio}
\end{figure}

% -------------------------------------------------------------------------------------
\subsection{Generalized virial equations and virial equilibria}
\label{sec:ve2}

Multiplying both sides of the velocity-moment equations~(\ref{eq:meq2}) by $4\pi r^3$ and integrating over the radius $r$, we can obtain the generalized virialization relations as
%-----------------------------------------------------------------------------------
% integral form
\begin{displaymath}
- 2 (m^2-1)T_{k+2,m} + 2 (k+1)(m+2) T_{k,m+2}
\end{displaymath}
\begin{equation}
\label{eq:exvir}
+(k+1)(m+1)W_{k,m} = 4\pi (m+1) r^3 P_{k+2,m},
\end{equation}
in which
% -------------------------------------------------------------------------------------
%Definition of T, W
% generalized virilization relation
\begin{eqnarray}
\label{eq:gvir}
T_{k,m} & \equiv & \frac{1}{2}\int^r_0 \rho\overline{v^k_r v^m_{t}} 4\pi r^2\dd r, \nonumber \\
W_{k,m} & \equiv & -\int^r_0 \rho\overline{v^k_r v^m_{t}}\frac{\dd\Phi}{\dd r} 4\pi r^3 \dd r
\end{eqnarray}
and $P_{k,m} \equiv \rho\overline{v^k_r v^m_t}$. When $k = m = 0$, equation~(\ref{eq:exvir}) reduces to
% -------------------------------------------------------------------------------------
\begin{equation}
\label{eq:vir1}
2 k_{\rm tot}(r) + u(r) = 4\pi r^3 p_r(r),
\end{equation}
where $k_{\rm tot}(r) = T_{2,0} + 2 T_{0,2}$ and $u (r) = W_{0,0}$ are the total kinetic and potential energy contained in the $r$-sphere, and $p_r(r) = P_{2,0}$ is the radial pressure. By differentiating equation~(\ref{eq:exvir}) with respect to $r$, we can immediately restore the moment equations~(\ref{eq:meq2}). It can be seen that the moment equations are equivalent to the generalized virial equations~(\ref{eq:exvir}), and hence the mechanical equilibria are identical to the virial equilibria.

From equation~(\ref{eq:vir1}), we define the virialization ratio $r_v$ as a function of $r$ as
% -----------------------------------------------------------------------------------
% definition of virialization ratio
\begin{equation}
\label{eq:rv}
r_v(r) \equiv \frac{2 k_{\rm tot}(r)}{-u(r) + 4\pi r^3 p_r(r)}.
\end{equation}
With the density profile $\rho(r)$, the velocity dispersion profiles $\sigma^2_r(r)$ and $\sigma^2_t(r)$ of the simulated dark matter halo \citep[see][]{navarro10,ludlow10}, we calculate $r_v$, as shown in Fig.~\ref{fig:vratio}. It can be seen that $r_v$ nearly equals one at all the radii, except for the inner fluctuation and a peculiarity at $r \approx 100 {\rm kpc/h}$. This agreement indicates that the simulated dark halo is well virialized, and it also implies that the CBE is a very good approximation for describing self-gravitating systems, such as dark matter haloes.

%--------------------------------------------------------------------------------------
\begin{figure}
\centerline{\includegraphics[width=0.8\columnwidth]{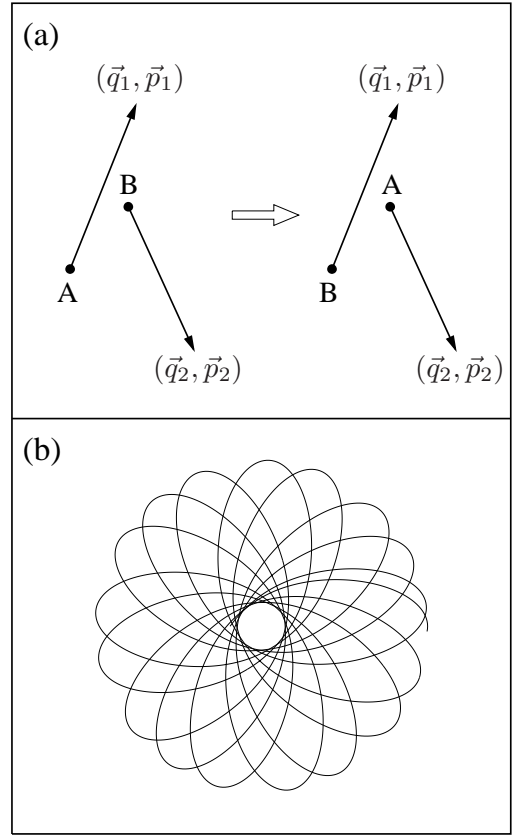}}
\caption{Illustration of the ergodicity validity for short-range interaction systems and the global ergodicity breaking for self-gravitating systems. Panel~(a) shows the microstates of a short-range interaction system (e.g. an ideal gas), in which the motion states of two particles, as a result of stochastic motions driven by irregular forces, can be exchanged with each other. Panel~(b) shows the typical non-resonant regular orbit of a gravitating particle in the static spherical potential of the self-gravitating system, in which the global ergodicity is invalid.}
\label{fig:ergodic}
\end{figure}

% -------------------------------------------------------------------------------------
\begin{figure*}
\centerline{\includegraphics[width=1.75\columnwidth]{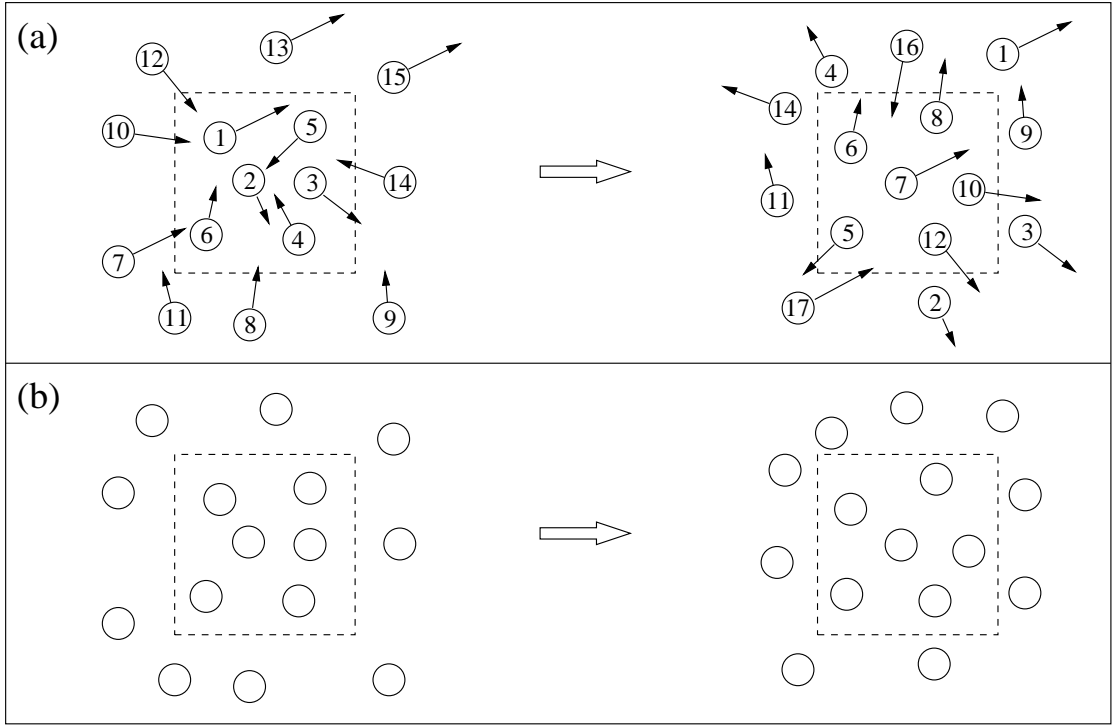}}
\caption{Illustration of the re-establishment of the local ergodicity for collisionless equilibrated self-gravitating systems. For simplicity, the motion states at only two instants are shown. (a) Within a finitely small volume element of the system, depicted by the fictitious dashed square, all the collisionless gravitating particles have deterministic trajectories and labels, and are moving extremely rapidly in randomly oriented straight lines. At the next instant, the original particles move out of the volume element, and are superseded by some other particles from the outside. (b) If, within this very volume element, these randomly and homogeneously distributed particles are treated as indistinguishable, they will exhibit a static distribution pattern. These patterns at different instants constitute an ensemble, every member of which occurs with equiprobability in the hypersurface, defined by all the invariant velocity moments, for the volume element. In this sense, the microstates of the element (i.e. the local microstates) can be defined, and the local ergodicity can be re-established.}
\label{fig:indistinguish}
\end{figure*}

%---------------------------------------------------------------------------------------
\section{Ergodicity global breaking and local validity}
\label{sec:ergodic}

Ergodicity, or equiprobability principle, is the fundamental for equilibrium statistical mechanics. In physics and thermodynamics, the ergodic hypothesis says that, over long periods of time, the time spent by a particle in some region of the phase space of microstates with the same energy is proportional to the volume of this region (i.e. that all accessible microstates are equiprobable over a long period of time)\footnote{This statement of ergodic hypothesis is taken from Wikipedia at http://en.wikipedia.org/wiki/Ergodic\_hypothesis.}. Below, we illustrate that the ergodic hypothesis is not valid for the whole self-gravitating system, but can be re-established in any local finitely small volume element.

%---------------------------------------------------------------------------------------
% global ergodic
\subsection{Complete relaxation and ergodicity validity}
\label{sec:cergodic}

In the usual thermodynamic system with short-range interactions, such as an ideal gas,
when ergodicity is valid, the constituent particles are driven by irregular forces and perform Brownian motions. As a result, the particles can traverse any place of the system, and can attain any possible value of the momentum (or velocity). In this sense, the stochastic motion of particles and the complete relaxation of the system are, respectively, the necessary and sufficient condition for the validity of ergodicity. For instance, in an ideal gas, the motion states at some instant $t_1$ of the two particles $A$ and $B$ in Fig.~\ref{fig:ergodic}(a) are $(\vc{q}_1, \vc{p}_1)$ and $(\vc{q}_2, \vc{p}_2)$, respectively. Because of the randomness of the motion, at some later time $t_2$, the microstate of the system is also available by exchanging the motion states between $A$ and $B$. If the two microstates before and after the exchanging of the two particles are equiprobable, i.e. $P[A(\vc{q}_1, \vc{p}_1), B(\vc{q}_2, \vc{ p}_2)] = P[A(\vc{q}_2, \vc{p}_2),B(\vc{q}_1, \vc{p}_1)]$, then the ergodicity is valid, and vice versa.

%---------------------------------------------------------------------------------------
% global ergodic
\subsection{Incomplete relaxation and ergodicity global breaking}
\label{sec:gergodic}

However, the above statement of ergodicity is invalid for the whole collisionless self-gravitating system. Fig.~\ref{fig:ergodic}(b) shows the typical regular, or non-resonant, orbit of a gravitating particle moving in an equilibrated spherical gravitational field. An unclosed curve resembles a rosette in an orbital plane, and eventually passes close to every point in the annulus of an invariant torus, on which the energy and angular momentum of the particle are conserved \citep{galdyn08}. Even if the orbit suffers from some small perturbations, according to the Kolmogorov - Arnold - Moser theorem, the torus corresponding to the regular orbit with fundamental frequencies sufficiently incommensurable will survive the small perturbations. It retains its topology such that the motion in its vicinity remains regular and confined to the slightly deformed invariant torus \citep{mo10}. Hence, such regular motion of the particle is definitely not ergodic motion, in which the particle cannot traverse any place of the system, and also cannot attain any value of the velocity \citep[see][]{hao04}.

% -------------------------------------------------------------------------------------
\begin{figure*}
\centerline{\includegraphics[width=1.75\columnwidth]{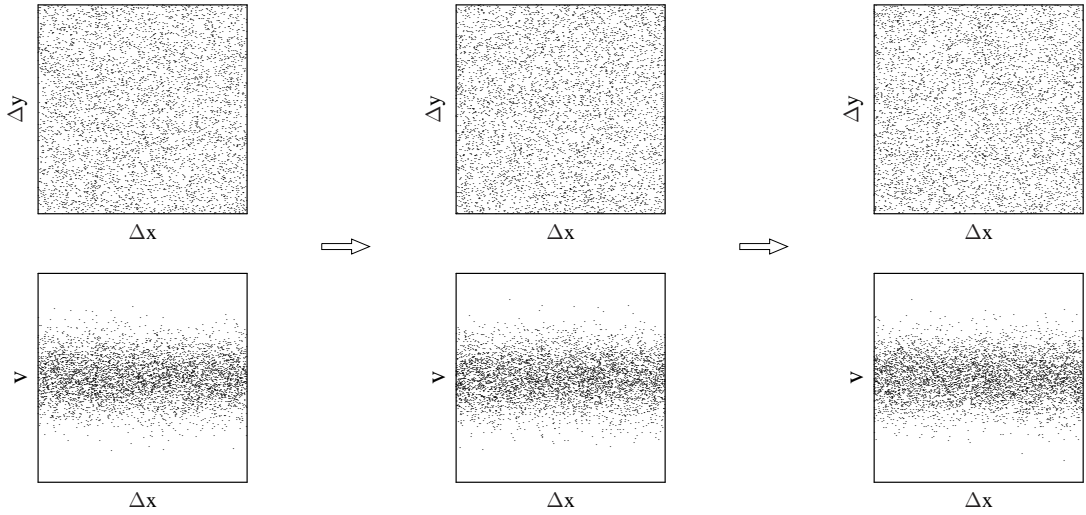}}
\caption{Illustration of the local ergodicity of indistinguishable particles for collisionless equilibrated self-gravitating systems. The upper three scatter plots are 2-dimensional slices of spatial distributions of the indistinguishable particles within a finitely small volume element. The lower plots are the corresponding 1-dimensional velocity distributions within this very element. The local ergodicity is valid, provided that the gravitating particles are treated as indistinguishable, that these particles are randomly and homogenously distributed within the finite volume element under consideration, and that the velocity distribution is independent of the spatial coordinates. The arrows indicate the time evolution of these distributions for this volume element. We see that, if the possible fluctuations are neglected, both the spatial and velocity distribution remain unchanged with time when the local ergodicity is established and the system is in equilibrium states.}
\label{fig:distxv}
\end{figure*}

% --------------------------------------------------------------------------------------
Additionally, \citet{vanalbada82} indicates that violent relaxation cannot wipe out all the information concerning the initial conditions when the system settles down into an equilibrium configuration (see their fig. 3). This correlation between the initial and final states also implies that violent relaxation cannot give rise to the global ergodicity for the whole system and hence the relaxation is incomplete.

\citet{lb67} derived a DF of self-gravitating systems, on the basis of the complete relaxation and global validity of ergodicity. However, the Lynden-Bell distribution leads to the isothermal mass profile of a gravitating system, with infinite mass, energy and spatial extent. \citet{lb67}, \citet{shu78} and \citet{madsen87} have already suggested that some kind of incomplete relaxation mechanism should be considered. For example, the suggested cure by \citet{lb67} for incomplete relaxation is to introduce a cut-off in phase-space, so that complete violent relaxation takes place in a limited region only. Shu's approach, however, is to consider additional macroscopic constraints \citep{shu69, shu87}.

Self-gravitating systems are not the only long-range interaction systems in which ergodicity is broken. \citet{campa09} have reported that many ergodicity-breaking events can be found in other long-range systems, such as the generalized Hamiltonian mean field model, the Ising and the $XY$ long- plus short-range model (see also \citealt{bouchet08}). So, it seems that the ergodicity breaking could be a common feature of long-range interaction systems.

% -------------------------------------------------------------------------------------
% local ergodicity
\subsection{Incomplete relaxation and ergodicity local validity}
\label{sec:lergodic}

Ergodicity is the basis of equilibrium statistical mechanics for any thermodynamic system. If ergodicity is completely invalid, then the equilibrium statistical mechanics cannot exist. Fortunately, we can re-establish the local ergodicity for equilibrated self-gravitating systems in the following way.

Imagine that, at some instant, within a finitely small (not infinitesimal) volume element of the system, all the collisionless gravitating particles have deterministic trajectories and labels, and are moving extremely rapidly in randomly oriented straight lines, as shown in Fig.~\ref{fig:indistinguish}(a). At the next instant, some of the original particles move out of the volume element, but some others will move into this very element. As time passes, this process occurs repeatedly and continuously, so that the particles are constantly superseded. Because these particles are not confined within this volume element, if the observer's field of vision is restricted within this element, these particles cannot trackable. In this case, it is not possible to define the microstates of the local element. However, if these particles are treated as indistinguishable, even if they have definite trajectories and labels, then the physical picture of the microscopic motions is renewed, as shown in Fig.~\ref{fig:indistinguish}(b). In this picture, at every instant, these particles, which are losing their trajectories and labels, are randomly and homogeneously populated in the volume element under consideration. The homogeneous distribution of these indistinguishable particles at an instant exhibits a static distribution pattern, and these different patterns at different instants naturally constitute an ensemble. When the system is in its equilibrium state, all the background field quantities (i.e. the velocity moments given by equations~\ref{eq:vm} or \ref{eq:vmts}) fluctuate at a magnitude proportional to the inverse of $\sqrt{\Delta N}$, where $\Delta N$ is the particle number of this volume element, with $\Delta N \gg 1$. So, these moments can be considered as local invariants, which define a confined hypersurface for this volume element. It is reasonable to assume that every member of the ensemble of the local distribution patterns occurs with equal probability in this hypersurface. So, in this sense, the microstates of the volume element (i.e., the local microstates) can be defined, and hence the local ergodicity can be re-established.

We provide an intuitive understanding of how to identify the local ergodicity. If the particles are treated as indistinguishable, if these particles are randomly and homogenously distributed within the finite volume element under consideration and if the velocity distribution is independent of the spatial coordinates, then the local ergodicity can be re-established (see Fig.~\ref{fig:distxv}). Such a requirement is extremely easy to satisfy, and hence the local ergodicity should always be valid.

In the previous subsection, we proved the invalidity of global ergodicity by regarding the gravitating particles as distinguishable, with deterministic trajectories and labels. Even if the particles are globally treated as indistinguishable, the validity of global ergodicity cannot be recovered. According to the ensemble theory, the probability distribution of the microstates of a closed system at a fixed energy is derived by assuming that all states on the energy hypersurface in phase space have equal probability. However, as \citet{galdyn08} have argued, the hypersurface of a constant energy for an isolated self-gravitating system is unbounded, and thus the microcanonical probability distribution cannot be defined. This implies that ergodicity, or equiprobability, is broken globally.

%-------------------------------------------------------------------------------------
\begin{figure*}
\centerline{\includegraphics[width=1.75\columnwidth]{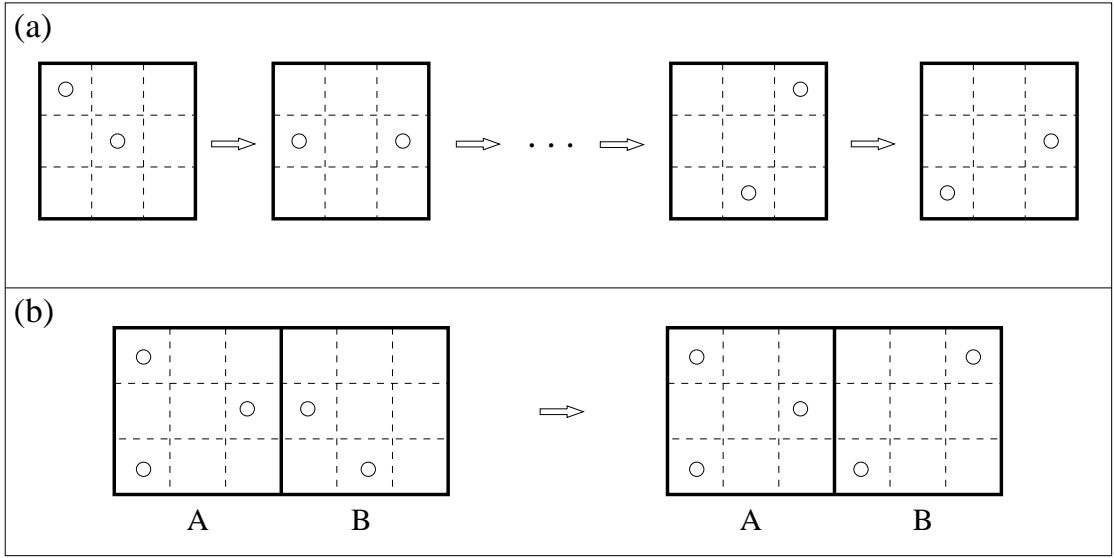}}
\caption{(a) Illustration of the Fermi-Dirac-like statistics for self-gravitating systems. The local finitely small phase-space element (i.e. the macrocell) is divided into many microcells (nine in the figure). The possible minimum phase-space volume is $\Delta x \Delta v = h_g$, whose value is small enough so that each microcell can accommodate only one gravitating particle (as it is impossible for two gravitating particles to be close to each other but simultaneously to be relatively static). However, the value is large enough that there are no two-body encounters between particles in adjacent microcells. Hence, the microstates of the indistinguishable gravitating particles within this macrocell should be counted by Fermi-Dirac-like statistics. Our approach is more like the particle approach of \citet{shu78}, but such an exclusion effect was already noticed by \citet{lb67}, when he designed his `fourth statistics'. We assume that there are two (indistinguishable) particles. Because of the local ergodicity, the two particles will populate these microcells with equal probability, which also implies that all the microstates for this macrocell occur with equal probability. According to the Fermi-Dirac-like statistics, there are in total $36$ microstates for the two indistinguishable particles in the macrocell (only four of them are shown). (b) Illustration of the spatial independence of the microstates. Here, $A$ and $B$ are two macrocells that are spatially adjacent. Because both the direct gravitational interaction and the statistical correlation between any two gravitating particles can be neglected, which is exactly the same reason to justify the CBE, it is reasonable to assert that the microstates of the two particles in $B$ do not depend on the motion states of the three particles in $A$. That is, the distribution patterns of the two particles at the two instants in $B$ occur with equiprobability, and hence the spatial independence of the local microstates can be established.}
\label{fig:macrocell}
\end{figure*}

%----------------------------------------------------------------------------------------
% ensemble inequivalence
\subsection{Negative specific heat and ensemble inequivalence}
\label{sec:enineq}

The paradox of ensemble inequivalence comes from the following argument \citep[see][]{lb68, thirring70, thirring71, campa09}. In the canonical ensemble, the heat capacity at constant volume is
% --------------------------------------------------------------------------------------
\begin{equation}
\label{eq:cv1}
C_V = \frac{\partial \langle E \rangle}{\partial T} = \beta^2 \langle(E-\langle E\rangle)^2\rangle > 0,
\end{equation}
where $\langle E \rangle$ is the mean value of the energy. Whereas for the self-gravitating system at constant energy (i.e. in the microcanonical ensemble), according to the virial theorem, $\langle U \rangle = -2\langle K \rangle$, so $E = \langle K \rangle + \langle U \rangle = - \langle K \rangle$. Because the kinetic energy defines the temperature, as $T \propto K$, we obtain the heat capacity of the microcanonical ensemble as
% --------------------------------------------------------------------------------------
\begin{equation}
\label{eq:cv2}
C_V = \frac{\partial E}{\partial T} \propto \frac{\partial E}{\partial K} < 0.
\end{equation}
With the positive and negative specific heat derived in equations~(\ref{eq:cv1}) and (\ref{eq:cv2}), respectively, we can see that the microcanonical ensemble and the canonical ensemble are inequivalent for self-gravitating systems.

We propose that such a paradox should originate from the misuse of the ensemble theory. In fact, a detailed analysis indicates that the ensemble theory is established only upon the validity of global ergodicity, and hence is only applicable for short-range interaction systems. In Section~\ref{sec:gergodic}, we have demonstrated that the global ergodicity is not valid for self-gravitating systems, so the relationships of equations~(\ref{eq:cv1}) and (\ref{eq:cv2}), and any other conclusions based on ensemble theories, are rootless for self-gravitating systems \citep[e.g.][]{chavanis05, chavanis06}.

As mentioned in Section~\ref{sec:lergodic}, \citet{galdyn08} have presented an alternative argument against the existence of the microcanonical ensemble for self-gravitating systems. They have suggested that the hypersurface of a fixed energy for an isolated self-gravitating system is unbounded, and thus the microcanonical probability distribution cannot be defined. Moreover, they have also demonstrated that the canonical ensemble does not exist. Hence, without the validity of ensemble theory, the relationships of equations~(\ref{eq:cv1}) and (\ref{eq:cv2}) should be, at most, considered as two different definitions of the so-called `heat capacity' for self-gravitating systems.

In the next section, we demonstrate the significance of this local ergodicity in deriving the Boltzmann-Gibbs entropy for self-gravitating systems.

% --------------------------------------------------------------------------------------
\section{Boltzmann-Gibbs/White-Narayan statistics}
\label{sec:core}

As addressed in the introduction, in the previous studies \citep{hep10,kang11}, we found the following. (i) Both the concept of entropy and the entropy principle are still valid for self-gravitating systems. (ii) Entropy is additive and hence extensive. (iii) The equilibrium states consist of mechanical equilibria. (iv) The local statistical properties differ from the global properties. Indeed, we have demonstrated in Section~\ref{sec:gergodic} that the global ergodicity for self-gravitating systems is broken. However, if the particles are treated as indistinguishable, then the local ergodicity can still be re-established. Such a characteristic of ergodicity suggests that the statistical mechanics of self-gravitating systems should be quite different from the standard statistical mechanics of short-range interaction systems.

% ------------------------------------------------------------------------------
% Boltzmann-Gibbs S
\subsection{Boltzmann-Gibbs entropy and the coarse-grained DF}
\label{sec:bgs}

We first derive the appropriate entropy form for self-gravitating systems, on the basis of the local ergodicity and the statistical independence of particle positions suggested by equation~(\ref{eq:nbdf}).

The local ergodicity implies that the indistinguishable gravitating particles should obey Fermi-Dirac-like statistics (see Fig.~\ref{fig:macrocell}a). We assume that the center of a finite volume element is located at $\vc{x} = (x_1, x_2, x_3)$, whose volume is $\Delta^3\vc{x}$. The particle number in this element is $\Delta N(\vc{x}) = \rho_N(\vc{x})\Delta^3\vc{x}$, where $\rho_N$ is the particle number density, a function of $\vc{x}$. By specifying the particle mass, $m_{\rm p}$, we can convert $\rho_N$ to mass density $\rho_m$ as $\rho_m = m_{\rm p}\rho_N$. Let $n(\vc{x}, \vc{v})$ denote the velocity distribution, which is the particle number in the finitely small phase-space element (i.e. the macrocell), $\Delta^3\vc{ x} \Delta^3\vc{v}$, located at the phase point $(\vc{x}, \vc{v})$. According to the definition of the coarse-grained DF $f_c(\vc{x}, \vc{v})$ in equation~(\ref{eq:cgdf1}), we have $n(\vc{x}, \vc{v}) \propto f_c(\vc{x}, \vc{v})\Delta^3\vc{x} \Delta^3\vc{v}$, and hence $\Delta N(\vc{x}) = \sum_{\vc{v}} n(\vc{x}, \vc{v})$. As analyzed above, the distribution $n (\vc{x}, \textbf {\emph v})$ should obey Fermi-Dirac-like statistics. So, the number of the microstates in this volume element is
% ------------------------------------------------------------------------------------
\begin{displaymath}
w[n,\vc{x}]=\prod\limits_{\vc{v}}\frac{\omega(\vc{x}, \vc{v})!} {n(\vc{ x}, \vc{v})! (\omega(\vc{x}, \vc{v}) - n(\vc{x}, \vc{v}))!} \approx\prod \limits_ {\vc{v}} \frac{\omega(\vc{x}, \vc{v})^{n({\scriptsize \vc{x}, \vc{v}})}}{n(\vc{x}, \vc{ v})!}
\end{displaymath}
%-------------------------------------------------------------------------------------
% W for gravitating particles
\begin{equation}
\label{eq:wgp}
=\frac{1}{\Delta N !}\left(\frac{\Delta N!} {\prod\limits_{\vc{v}} n(\vc{x}, \vc{v})!} \prod\limits_{\vc{v}} \omega(\vc{x}, \vc{v}) ^{n({\scriptsize \vc{x}, \vc{v}})} \right) = \frac{w_{\rm MB}[n,\vc{x}]}{\Delta N !},
\end{equation}
where $\omega(\vc{x}, \vc{v})=\Delta^3\vc{x} \Delta^3 \vc{v}/h^3_g$ denotes the degeneracy at the velocity level $\vc{v}$ (not energy level) for the macrocell. If, for all velocities $\vc{v}$, $n(\vc{x}, \vc{v})/\omega(\vc{x}, \vc{v}) \ll 1$ (i.e. if the distribution satisfies the non-degenerate condition), we obtain the approximated equality in the first line of equation~(\ref{eq:wgp}). Here, $w_{\rm MB}[n, \vc{x}]$ denotes the microstate number counted with the Maxwell-Boltzmann statistics, given by the term enclosed in the parentheses.

Because the close encounters between gravitating particles can be neglected, which is the reason for justifying the CBE, in this case $h_g$ can be treated as an infinitesimal quantity (i.e. $h_g \rightarrow 0$), and hence $\omega \rightarrow \infty$. As a result, the validity of the CBE is consistent with the non-degenerate condition, $n/\omega \ll 1$. Whereas in very dense regions, such as the center of the dark matter halo, where two-body relaxation might play an important role, the CBE might not be a good approximation. In this case, the non-degenerate condition will be invalid.

As illustrated in Fig.~\ref{fig:macrocell}(b), it is reasonable to believe that the microstates of different macrocells are spatially independent of each other. Hence, the total number of the microstates in the whole system is constructed as the product of $w[n, \vc{x}]$ of all the volume elements,
% -------------------------------------------------------------------------------------
% total W for gravitating particles
\begin{equation}
\label{eq:twgp}
W[n]=\prod\limits_{\vc{x}}w[n,\vc{x}],
\end{equation}
where $w[n, \vc{x}]$ and $W[n]$ are both functionals of the distribution $n(\vc{x}, \vc{v})$. Taking the logarithm of both sides of equation~(\ref{eq:twgp}), we have
% ----------------------------------------------------------------------------------
\begin{displaymath}
\ln W[n] = -\sum_{\vc{x}, \vc{v}}\ln n(\vc{x}, \vc{v})! + \sum_{\vc{ x},\vc{v}} n(\vc{x}, \vc{v})\ln \omega(\vc{x}, \vc{v})
\end{displaymath}
\begin{equation}
\label{eq:logwn}
\approx N_{\rm T} - \sum_{\vc{x},\vc{v}} n(\vc{x}, \vc{v}) \ln n(\vc{x}, \vc{ v}) + \sum_{\vc{x}, \vc{v}} n(\vc{x}, \vc{v})\ln \omega(\vc{x}, \vc{v}).
\end{equation}
Here, we use Stirling's formula, $\ln n! \approx n(\ln n-1)$, for the approximation in the second line, and $N_{\rm T}$ is the total particle number of the system:
\begin{displaymath}
N_{\rm T}=\sum_{\vc{x}}\Delta N (\vc{x}) = \sum_ {\vc{x}, \vc{v}} n(\vc{x}, \vc{v}).
\end{displaymath}
By dropping the unimportant constant $N_{\rm T}$, with
% ------------------------------------------------------------------------------------
% coarse-grained DF-2
\begin{equation}
\label{eq:cgdf2}
F (\vc{x}, \vc{v}) \equiv \frac{n(\vc{x},\vc{v})} {\omega (\vc{x}, \vc{v})} = \frac{h^3_g}{\Delta^3\vc{x} \Delta^3 \vc{v}} n(\vc{x},\vc{v}) \propto f_c(\vc{ x}, \vc{v}),
\end{equation}
we obtain
% --------------------------------------------------------------------------------------
\begin{displaymath}
\ln W[F] = - \sum_{\vc{x}, \vc{v}} n(\vc{x}, \vc{v}) \ln n (\vc{x}, \vc{ v}) + \sum_{\vc{x}, \vc{v}} n(\vc{x}, \vc{v})\ln \omega(\vc{x}, \vc{v})
\end{displaymath}
\begin{displaymath}
= -\sum_{\vc{x}, \vc{v}}\omega(\vc{x}, \vc{v})\frac {n(\textbf {\emph x},\vc{ v})}{\omega(\vc{x},\vc{v})} \ln \frac{n( \vc{x}, \vc{v})} {\omega(\vc{x}, \vc{ v})}
\end{displaymath}
\begin{displaymath}
= - \frac{1}{h^3_g} \sum_{\vc{x}, \vc{v}} \Delta^3\vc{x} \Delta^3 \vc{v} F(\vc{ x}, \vc{v}) \ln F(\vc{x}, \vc{v})
\end{displaymath}
\begin{equation}
\label{eq:logwf}
= - \frac{1}{h^3_g}\int F(\vc{x}, \vc{v}) \ln F(\vc{x}, \vc{ v})\dd^3\vc{x} \dd^3\vc{v}.
\end{equation}
In the last line, $\sum \Delta^3 \vc{x} \Delta^3 \vc{v}$ is replaced by $\int\dd^3 \vc{x} \dd^3 \vc{v}$. Ignoring the unimportant factor $1/h^3_g$, we use the Boltzmann relation to define the entropy, $S\equiv\ln W$, and then the total entropy of the system is exactly the Boltzmann-Gibbs entropy:
% --------------------------------------------------------------------------------------
% Boltzmann-Gibbs entropy density
\begin{equation}
\label{eq:bgs}
S[F] = -\int F(\vc{x}, \vc{v})\ln F(\vc{x}, \vc{v}) \dd^3\vc{x}\dd^3 \vc{v}.
\end{equation}
As mentioned previously, the entropy for self-gravitating systems should be additive and hence extensive. We can see that the extensivity of entropy comes from the spatial independence of microstates (equation~\ref{eq:twgp}). As in standard statistical mechanics, the factor $1/\Delta N!$ in equation~(\ref{eq:wgp}) ensures the extensivity of the entropy of equation~(\ref{eq:logwn}) or equation~(\ref{eq:bgs}).

From equation~(\ref{eq:cgdf2}), we see that the DF $F(\vc{x}, \vc{v})$ is the same as $f_c(\vc{x}, \vc{v})$ except for a constant coefficient. Thus, we can regard them as identical. The DF $F(\vc{x}, \vc{v})$ can be expressed by Taylor expansion with respect to $\vc{v}$, as
% ------------------------------------------------------------------------------------
\begin{equation}
\label{eq:taylor}
\ln F(\vc{x}, \vc{v}) = \sum_{k,m,n} \frac{1}{k!m!n!} \frac{\partial^{k+m+n}\ln F(\vc{x}, \vc{v})} {\partial v^k_1 \partial v^m_2 \partial v^n_3}\big|_{\vc{v}=0} v^k_1 v^m_2 v^n_3.
\end{equation}
So, we have
% ------------------------------------------------------------------------------------
\begin{equation}
\label{eq:lmpd2}
F(\vc{x}, \vc{v}) = \exp\left(-\sum_{k,m,n} \lambda_{k,m,n} (\vc{x}) v^k_1 v^m_2 v^n_3\right),
\end{equation}
with
% -----------------------------------------------------------------------------------
% Taylor expansion
\begin{equation}
\label{eq:lagtlor}
\lambda_{k,m,n} (\vc{x}) = -\frac{1}{k!m!n!} \frac{\partial^{k+m+n} \ln F(\vc{ x}, \vc{v})} {\partial v^k_1 \partial v^m_2 \partial v^n_3}\big|_{\vc{v}=0},
\end{equation}
where $\lambda_{k,m,n} (\vc{x})$ are functions of the spatial coordinates $\vc{ x}$. We should emphasis that the fine-grained DF $f(\vc{x}, \vc{v})$ obeys the Jeans theorem that, as a steady-state solution of the CBE, $f(\vc{x}, \vc{v})$ depends on the phase-space coordinates only through integrals of motion in the given potential, such as the Hamiltonian $H$, or angular momentum $\vc{L}$. However, the coarse-grained DF $F(\vc{x}, \vc{v})$ does not obey the CBE, and hence it is not subject to a constraint of the Jeans theorem.

The formulation of $F(\vc{x}, \vc{v})$, which we have adopted in equation~(\ref{eq:lmpd2}), is the fundamental expression. However, the formulation is not unique, and we can choose various alternative forms to express $F(\vc{x}, \vc{v})$. With the fundamental expression in equation~(\ref{eq:lmpd2}), we can define a quantity, $Z_g(\vc{x})$, as
% -------------------------------------------------------------------------------------
% partition function
\begin{equation}
\label{eq:zf}
Z_g(\vc{x}) \equiv \int F_{-0}(\vc{x}, \vc{v}) \dd^3 \vc{v},
\end{equation}
where $F_{-0}(\vc{x}, \vc{v})$ is defined by removing the zero-order term from $F(\vc{x}, \vc{v})$, as $F_{-0} (\vc{x}, \textbf{ \emph v}) \equiv F(\vc{x}, \vc{v})/\exp (-\lambda_{0,0,0})$. The non-vanishing velocity moments of various orders, $M_{k,m,n} (\vc{x})$, relate to the coarse-grained DF as
% --------------------------------------------------------------------------------------
% velocity moments (vmts)
\begin{displaymath}
\label{eq:vmts}
M_{k,m,n}(\vc{x}) \equiv \rho\overline{v^k_1 v^m_2 v^n_3}(\vc{x}) \equiv\int v^k_1 v^m_2 v^n_3 F(\vc{x},\vc{v})\dd^3 \vc{v}
\end{displaymath}
\begin{displaymath}
= \exp(-\lambda_{0,0,0}) \int v^k_1 v^m_2 v^n_3 F_{-0}(\vc{x}, \vc{v}) \dd^3 \vc{v}
\end{displaymath}
\begin{equation}
\label{eq:vmts}
= -\exp(-\lambda_{0,0,0})\frac{\dd Z_g(\vc{x})}{\dd\lambda_{k,m,n}} = -\rho (\vc{x})\frac{\dd \ln Z_g(\vc{x})}{\dd\lambda_{k,m,n}}.
\end{equation}
Here, $k,m,n$ are non-negative integers (i.e. $k, m, n = 0, 1, 2, \ldots$), $\rho (\vc{x})$ is the mass or number density, and we have used
\begin{displaymath}
\exp(-\lambda_{0,0,0}) = \frac{M_{0,0,0}(\vc{x})}{Z_g(\vc{x})} = \frac{\rho(\vc{x})} {Z_g(\vc{x})}.
\end{displaymath}
As explained in Section~\ref{sec:me2}, the coarse-grained moments differ from the fine-grained moments by only up to second-order small quantities, determined by the size of the finite volume element. So, the moments evaluated by equation~(\ref{eq:vmts}) can be directly substituted into the moment equations~(\ref{eq:meq2}).

We can see that the function $Z_g(\vc{x})$ greatly resembles the partition function of the usual statistical mechanics of short-range systems. Here, we also call $Z_g(\vc{x})$ the partition function -- of the self-gravitating systems.

From equation~(\ref{eq:vmts}), we can see that any moment can be derived from the partition function, $Z_g(\vc{x})$, by differentiating with respect to the corresponding coefficient of Taylor expansion. However, it would be a great mathematical challenge to evaluate simultaneously both the partition function $Z_g(\vc{x})$ and the entropy $S[F]$ of equation~(\ref{eq:bgs}). We leave these mathematical problems for future work.

To summarize, we identify the local ergodicity for self-gravitating systems, and we rederive the Boltzmann-Gibbs entropy by using Fermi-Dirac-like statistics, the non-degenerate condition and the spatial independence of local microstates. The Boltzmann-Gibbs entropy is confirmed to be suitable for the statistical mechanics of self-gravitating systems.

We emphasize once more that Binney's argument cannot invalidate the Boltzmann-Gibbs entropy for self-gravitating systems, because we have already revealed that equilibrated self-gravitating systems are not in maximum entropy states \citep[see][]{hep10,hep11}.

% ------------------------------------------------------------------------------------
\subsection{Macroscopic constraints and extended White-Narayan statistics}
\label{sec:macwn}

As mentioned in the introduction, the final relaxed state of self-gravitating systems is also the statistical equilibrium state. Therefore, it is the task for the statistical mechanics to completely determine these expansion coefficients in equation~(\ref{eq:lmpd2}).

It is easy to think that mass, energy, momentum and angular momentum conservation are possible macroscopic constraints for constructing the statistical mechanics of self-gravitating systems. If the DF is an even function of velocity ${\vc v}$ and the system is spherically symmetric, then the momentum and angular momentum conservation are automatically satisfied, in that they are both vanishing. Here, we do not consider the case of non-vanishing angular momentum. Besides these usual conservation constraints, however, the conservation of the phase-space volume of the fine-grained distribution should also be included \citep{lb67}. Moreover, as suggested by \citet{shu69, shu87}, the incomplete relaxation should be properly handled by introducing some macroscopic constraints. \citet{madsen87} has also suggested that the Vlasov equation should be properly considered.

We argue that, besides the usual constraints of the given mass and energy, the series of moment equations~(\ref{eq:meq2}) of the steady-state CBE must be included as additional constraints on the system's entropy functional when performing the variational calculus. Actually, the conservation of the phase-space volume of fine-grained distribution is a natural consequence of the CBE. As proved by \citep{cubarsi10}, the series of moment equations is equivalent to the CBE. So, if moment equations are included, the phase-space volume conservation is automatically satisfied.

However, it would be very inconvenient to calculate the variation using these moment equations directly. Like the example we will present in the next subsection with a truncated DF to the second-order expansion of velocity, we will use the equivalent series of generalized virialization relations (equations~\ref{eq:exvir}), instead of the original moment equations~(\ref{eq:meq2}), as the constraints of mechanical/virial equilibria. As a result, such a replacement will make the variational calculus more easy to perform.

Such an approach to the statistical mechanics for self-gravitating systems stems from the work of \citet{white87}, who applied the entropy of an ideal gas, $S_{\rm t} =\int^R_0 4\pi r^2 \rho \ln (p^{3/2}\rho^{-5/2})$, to gravitating systems, with the constrained entropy, as in the following (their equation~12):
% --------------------------------------------------------------------------------------
\begin{eqnarray}
\label{eq:wneq12}
Q & = & S_{\rm t} + \mu M_{\rm t} + \lambda E_{\rm t} + \nu p(R) \nonumber \\
  &   & + \int^R_0\left(\frac{\dd p}{\dd r}+\frac{GM\rho}{r^2}\right)f(r)4\pi r^2\dd r.
\end{eqnarray}
Here, $\nu p(R)$ represents the truncation at the radius $R$. Notice that, besides the usual constraints of mass and energy conservation (i.e. $\mu M_{\rm t}$ and $\lambda E_{\rm t}$). \cite{white87} also included the isotropic Jeans equation (i.e. the last integral term) as an additional constraint. By extremizing $Q$ with respect to $\rho$ and $p$, they derived an equation for the mass density of fully relaxed galaxies.

Although they did not obtain acceptable results, this approach still provides a correct line of thought, because, as we have mentioned previously, the final relaxed states of self-gravitating systems are not only statistical equilibrium states, but also mechanical equilibrium states. Hence, the mechanical equilibrium equations should also be considered as constraints to the total entropy $S_{\rm t}$, and the Jeans equation represents the second-order mode of the mechanical equilibria of self-gravitating systems. Below, we generalize the approach of \citet{white87} to formulate the methodology of the statistical mechanics for self-gravitating systems.

With the DF $F(\vc{x}, \vc{v})$ given by equation~(\ref{eq:lmpd2}), the total entropy from equation~(\ref{eq:bgs}) is
% --------------------------------------------------------------------------------------
% total Boltzmann-Gibbs entropy (tbgs)
\begin{eqnarray}
\label{eq:tbgs}
S_{\rm tot} & = & -\int F(\vc{x}, \vc{v}) \ln F(\vc{x}, \vc{v}) \dd^3\vc{x} \dd^3\vc{v} \nonumber \\
& = & \sum_{k,m,n}\int\lambda_{k,m,n} (\vc{x}) v^k_1 v^m_2 v^n_3 F(\vc{x}, \vc{ v}) \dd^3\vc{x} \dd^3\vc{v} \nonumber \\
& = & \sum_{k,m,n}\int\lambda_{k,m,n}(\vc{x})M_{k,m,n}(\vc{x}) \dd^3\vc{x},
\end{eqnarray}
where we used equation~(\ref{eq:vmts}) for the last equality. Because these moments, $ M_{k,m,n} (\vc{x})$, depend on the expansion coefficients, $\lambda_{k,m,n} (\vc{x})$, so the total entropy $S_{\rm tot}$ is a functional of $\lambda_ {k,m,n} (\vc{x})$.

From equation~(\ref{eq:vmts}), the mass density $\rho(\vc{x})$ and kinetic energy density $\epsilon_{\rm k}(\vc{x})$ are
% --------------------------------------------------------------------------------------
% mass and kinetic energy density
\begin{eqnarray}
\label{eq:mkdens}
\rho(\vc{x}) & = & M_{0,0,0}(\textbf {\emph x}), \nonumber \\
\epsilon_{\rm k}(\vc{x}) & = & \frac{1}{2}\rho(\vc{x}) (\overline{v^2_1}(\vc{ x}) + \overline{v^2_2}(\vc{x}) + \overline{v^2_3}(\vc{x})) \nonumber \\
& = & \frac{1}{2}(M_{2,0,0}(\textbf {\emph x}) + M_{0,2,0}(\textbf {\emph x}) + M_{0,0,2} (\textbf {\emph x})).
\end{eqnarray}
So the total mass, kinetic, and potential energy of the system are, respectively,
% -------------------------------------------------------------------------------------
\begin{eqnarray}
M_{\rm T} & = & \int\rho(\vc{x}) \dd^3 \vc{x}, \nonumber \\
E_{\rm K} & = & \int\epsilon_{\rm k}(\vc{x})\dd^3\vc{x}, \nonumber\\
E_{\rm V} & = & -\int\frac{G\rho(\vc{x})\rho(\vc{x}')}{2 | \textbf {\emph x} - \vc{x}'|} \dd^3\vc{x}\dd^3\vc{x}'.
\end{eqnarray}
Here, the integrations are performed over all the spatial volume, and hence the total energy is $E_{\rm T} = E_{\rm K} + E_{\rm V}$. As is the case for usual thermodynamic systems, in order to derive the proper thermodynamic relationships of self-gravitating systems using the entropy principle, the two common constraints of mass and energy conservation should be applied.

As addressed in Section~\ref{sec:me2}, the series of moment equations~(\ref{eq:meq2}) does not constitute a closed equation system, because there are no equations for $\rho\overline{v^k_t}$, the one-dimensional transverse moments of all orders ($k=0, 2, 4, \ldots$). This implies that $\lambda_ {k,m,n} (\vc{x})$, the coefficients of the Taylor expansion in equation~(\ref{eq:lmpd2}), cannot be completely determined (see Fig.~\ref{fig:moments}). It is the task of statistical mechanics to determine these coefficients completely. In Table~\ref{table1}, we summarize the numbers up to the 10th-order for both the moment equations and the expansion coefficients.

We have mentioned previously that mechanical equilibria must play crucial roles in establishing the statistical equilibrium states of self-gravitating systems. The moment equations~(\ref{eq:meq2}), or their equivalent virialization forms (equations~\ref{eq:exvir}), represent the mechanical/virial equilibria of various modes in spherically symmetric systems. We formally denote these virialization equations as non-spherical `${\rm Eqns}(\vc{x}, T, W, P) = 0$'. Therefore, the constrained entropy is
% -------------------------------------------------------------------------------------
% total constrained entropy
\begin{eqnarray}
\label{eq:tcs}
S_{\rm tot, c} & = & S_{\rm tot} + \mu_1 M_{\rm T} + \mu_2 E_{\rm T} + \nonumber \\
& & + \sum_{k,m,n} \int f_{k,m,n}(\vc{x}) {\rm Eqns}(\vc{x}, T, W, P) \dd^3 \vc{x},
\end{eqnarray}
where $\mu_1$, $\mu_2$ and $f_{k,m,n}(\vc{x})$ are Lagrangian multipliers. Compared with the original form of \citet{white87}, it can be seen that the mechanical/viral equilibria operate in very complicated ways, by including all orders of virialization equations into the constrained entropy. Because $S_{\rm tot}$, $M_{\rm T}$ and $E_{\rm T}$ are also functionals of the velocity moments, by performing the variational calculus, $\delta S_{\rm tot, c} = 0$, with respect to these moments, $M_{k,m,n}(\vc{x})$ [or equivalently $\lambda_{k,m,n} (\vc{x})$], we can formally obtain the differential equations concerning $M_{k,m,n} (\vc{x})$ [or $ \lambda_{k,m,n} (\vc{x})$], which characterize the statistical equilibria of the systems. These statistical equilibrium equations, together with the mechanical equilibrium equations, provide a complete description of the equilibrium states for self-gravitating systems.

Finally, we emphasis that we should not directly include the steady-state CBE into equation~(\ref{eq:tcs}) as the constraint in place of the series of virialization equations, because the entropy $S_{\rm tot}$ is constructed with the coarse-grained DF $F(\vc{x}, \vc{v})$, which does not satisfy the CBE. Moreover, there is no simple relation between $F(\vc{x}, \vc{v})$ and the fine-grained DF $f(\vc{x}, \vc{v})$.

We further explain the approach to the statistical mechanics that we have developed in the following example with a truncated DF.

% --------------------------------------------------------------------------------------
\begin{table}
\caption{Numbers of moment equations and Taylor expansion coefficients corresponding to the order of moments (see equations~\ref{eq:meq2} and \ref{eq:lmpd2}). Column~1 is the order of the velocity moments, column~2 is the number of moment equations of the order, and column~3 lists the cumulative number of the moment equations up to this order. Column~4 is the number of Taylor expansion coefficients of the order and column~5 is the cumulative number of expansion coefficients up to the order.}
\label{table1}
\begin{tabular}{ccccc}
\hline
Order   & Velocity  & Cumulative &   Taylor     & Cumulative    \\
 of     &  moment   &   moment   &  expansion   &  expansion    \\
moments & equations & equations  & coefficients & coefficients  \\ \hline
0       &      0    &     0      &       1      &       1       \\
2       &      1    &     1      &       2      &       3       \\
4       &      2    &     3      &       3      &       6       \\
6       &      3    &     6      &       4      &      10       \\
8       &      4    &    10      &       5      &      15       \\
10      &      5    &    15      &       6      &      21       \\ \hline
\end{tabular}
\end{table}

% -------------------------------------------------------------------------------------
% truncated DF
\subsection{Truncated DF of self-gravitating systems}
\label{sec:trundf}

For the spherically symmetric systems, we truncate the DF $F(\vc{x}, \vc{v})$ in equation~(\ref{eq:lmpd2}) to the second order, and we directly replace the expansion coefficients by the corresponding moments as
% ------------------------------------------------------------------------------------
% 3-D anisotropic DF
\begin{equation}
\label{eq:3df}
F(r, v_r, v_{\theta}, v_{\phi}) = \frac{\rho}{(2\pi)^{\frac{3}{2}}\sigma_r\sigma^2_{t}} {\rm exp} ( -\frac{v^2_r}{2 \sigma^2_r} - \frac{v^2_{\theta}}{2 \sigma^2_t} - \frac{v^2_{\phi}} {2 \sigma^2_t}).
\end{equation}
Here, the density $\rho$ and the velocity dispersions $\sigma^2_r$ and $\sigma^2_t$ are all functions of $r$. Because of the spherical symmetry, the dispersions of $v_{\theta}$ and $v_{\phi}$ are the same, so we uniformly use $\sigma^2_t$ to represent $\overline{v^2_{\theta}}$ or $\overline{v^2_{\phi}}$. As a result, the total (Boltzmann-Gibbs) entropy of the system with this truncated DF is
% -------------------------------------------------------------------------------------
% Boltzmann-Gibbs entropy
\begin{eqnarray}
\label{eq:bgt}
S_{\rm BG} & = & -\int F \ln F \dd^3\vc{x} \dd^3\vc{v} = \int^{\infty}_{0} \left[-\ln \left( \frac {\rho(r)} {\sigma_r(r) \sigma^2_t(r)} \right) \right. \nonumber\\
      & &+ \left.\frac{3}{2} (1 + \ln 2\pi)\right] \rho(r) 4\pi r^2 \dd r.
\end{eqnarray}
If we drop the unimportant constant term, $(3/2) (1 + \ln 2\pi)$, and extract the specific entropy from equation~(\ref{eq:bgt}), we have
% -------------------------------------------------------------------------------------
% Boltzmann-Gibbs specific entropy
\begin{equation}
\label{eq:bgss}
s = - \ln \left( \frac {\rho}{\sigma_r\sigma^2_t}\right) = \ln \left( \frac {p^{1/2}_r p_t}{\rho^{5/2}}\right),
\end{equation}
where $p_r \equiv \rho\sigma^2_r$ and $p_t \equiv \rho\sigma^2_t$ are the radial and one-dimensional tangential pressure, respectively. It can be seen that the specific entropy of equation~(\ref{eq:bgss}) greatly resembles that taken by \citet{white87}, except for the anisotropic velocity distribution (i.e. $p_r \neq p_t$).

The Jeans equation~(\ref{eq:2ndoe}) can be transformed into the virialization form of equation~(\ref{eq:vir1})
\begin{displaymath}
2k_{\rm tot}(r) + u(r) = 4\pi r^3 p_r(r),
\end{displaymath}
where $k_{\rm tot}(r)$ and $u(r)$ are the kinetic and potential energy contained in the $r$-sphere of the system. To be concise, $k_{\rm tot}(r)$ is expressed as
% -------------------------------------------------------------------------------------
% kinetic energy
\begin{equation}
\label{eq:ke}
k_{\rm tot}(r) = k_r(r) + 2 k_t(r),
\end{equation}
with (see equation~\ref{eq:gvir})
% --------------------------------------------------------------------------------------
% radial and tangential kinetic energy function
\begin{eqnarray}
\label{eq:rte}
k_r(r) &  = & T_{2,0} = \frac{1}{2}\int^r_0 4\pi r'^2 p_r(r')\dd r', \nonumber \\
k_t(r) &  = & T_{0,2} = \frac{1}{2}\int^r_0 4\pi r'^2 p_t(r')\dd r',
\end{eqnarray}
are the radial and one-dimensional tangential kinetic energy within the $r$-sphere, respectively. The potential energy $u(r)$ is expressed as
%----------------------------------------------------------------------------------
% potential energy function
\begin{eqnarray}
\label{eq:ue}
u(r) = W_{0,0} & = & -\int_{\Omega_r}\frac{G \rho(\vc{r}')\rho(\vc{r}'')}{2 |\vc{r}' - \vc{ r}''|}\dd^3 \vc{r}' \dd^3 \vc{r}'' \nonumber \\
    & = & -4 {\pi} G \int^r_{0}m(r')\rho(r')r' \dd r',
\end{eqnarray}
where the integral is confined within the volume of the $r$-sphere, $\Omega_r$, and $m(r)$ is the mass function
% -----------------------------------------------------------------------------------
% mass function
\begin{equation}
\label{eq:mf}
m(r) = \int^r_0 4 \pi r'^2 \rho(r')\dd r'.
\end{equation}

With these variable definitions, the total mass $M_{\rm T}$, the total kinetic energy $E_{\rm K}$ and the total potential energy $E_{\rm V}$ of the system are, respectively,
% --------------------------------------------------------------------------------------
% conserved quantities
\begin{equation}
\label{eq:consv}
M_{\rm T} = m(r)\Big|_{r \rightarrow \infty}, {\hskip 1.0mm} E_{\rm K} = k_{\rm tot}(r) \Big|_{r \rightarrow \infty}, {\hskip 1.0mm} E_{\rm V} = u(r)\Big|_{r \rightarrow \infty}.
\end{equation}

From equations~(\ref{eq:rte}) and (\ref{eq:mf}), we perform the following variable transformation by calculating the first derivatives as
% --------------------------------------------------------------------------------------
% variable-derivative-3
\begin{equation}
\label{eq:vd3}
\rho=\frac{m'}{4\pi r^2}, {\hskip 1cm} p_r=\frac{k'_r}{2\pi r^2}, {\hskip 1cm} p_t=\frac{k'_t}{2\pi r^2}.
\end{equation}
From equation~(\ref{eq:ue}), we have
% --------------------------------------------------------------------------------------
% potential energy-derivative
\begin{equation}
\label{eq:ub1}
u' = -4 {\pi} G m \rho r = -\frac{G m m'}{r}.
\end{equation}
The prime in equations~(\ref{eq:vd3}) and (\ref{eq:ub1}) indicates the first derivative with respect to the radius $r$.

Therefore, the total entropy of the system from equation~(\ref{eq:bgt}) is
% --------------------------------------------------------------------------------------
% total entropy -2
\begin{eqnarray}
\label{eq:tots2}
S_{\rm T} = \int^{\infty}_0 m'\left\{\ln \left[ \frac{k'^{1/2}_r (k'_{\rm tot} - k'_r)} {m'^{5/2}} \right] + \ln(2^{5/2} \pi r^2) \right\} \dd r.
\end{eqnarray}
With the variable transformations of equations~(\ref{eq:ke}), (\ref{eq:ue}) and (\ref{eq:mf}), the mass and energy conservation constraints for the variational calculus are converted in order to satisfy the fixed endpoint conditions of equation~(\ref{eq:consv}).

With equation~(\ref{eq:vd3}), equation~(\ref{eq:vir1}) can be re-expressed as
% -------------------------------------------------------------------------------
% virialization-2
\begin{equation}
\label{eq:vir2}
2 k_{\rm tot}(r) + u(r) - 2 r k'_r(r) = 0.
\end{equation}
We use this relation, instead of the original Jeans equation~(\ref{eq:2ndoe}), as the constraint of the mechanical equilibrium. As a result, such a transformation will make the variational calculus more easy to perform. Additionally, the mathematical relation between $u'$ and $m$, $m'$ in equation~(\ref{eq:ub1}), provides another additional constraint.

With all these constraints considered, the total constrained entropy of the system is
% ------------------------------------------------------------------------------------
\begin{displaymath}
S_{\rm T,c}=\int^{\infty}_0 H(r,m,m',k_{\rm tot},k'_{\rm tot},k_r,k'_r,u,u') \dd r
\end{displaymath}
\begin{displaymath}
{\hskip 7.5mm} = \int^{\infty}_0 \dd r \left\{m'\ln\left[\frac{k'^{1/2}_r (k'_{\rm tot} - k'_r)} {m'^{5/2}} \right] + m' \ln(2^{5/2} \pi r^2)\right.
\end{displaymath}
\begin{equation}
\label{eq:totcs}
{\hskip 10.9mm} + \left. f_1(r)(2k_{\rm tot} + u - 2 r k'_r) + f_2(r)(u' + \frac{G m m'}{r}) \right\}.
\end{equation}
Here, $H$ denotes the integrand (i.e. the terms enclosed in the braces) and $f_1$ and $f_2$ are two Lagrangian multipliers. Then, performing the standard variational calculus, $\delta S_{\rm T,c} = 0$, with respect to the variable pairs $(m, m'), (k_{\rm tot}, k'_{\rm tot}), (k_r, k'_r)$ and $(u, u')$, we obtain the following equations:
% ---------------------------------------------------------------------------------
\begin{equation}
\label{eq:steq1}
\frac{1}{2}\frac{\dd \ln p_r}{\dd r} + \frac{\dd \ln p_t}{\dd r} - \frac{5}{2}\frac{\dd \ln\rho}{\dd r} + f'_2\frac{G m}{r} = f_2 \frac{G m}{r^2},
\end{equation}
% ---------------------------------------------------------------------------------
\begin{equation}
\label{eq:steq2}
\frac{\dd}{\dd r} (\frac{\rho}{p_t}) = 2 f_1,
\end{equation}
% ---------------------------------------------------------------------------------
\begin{equation}
\label{eq:steq3}
\frac{\rho}{p_r} - \frac{\rho}{p_t} = 2 f_1 r + C_1,
\end{equation}
% ---------------------------------------------------------------------------------
\begin{equation}
\label{eq:steq4}
\frac{\dd f_2}{\dd r} = f_1.
\end{equation}
Here, again, the prime in equation~(\ref{eq:steq1}) indicates the first derivative with respect to $r$. In equation~(\ref{eq:steq3}), $C_1$ is an integration constant, but the end point of $k_r(r)$ at $r\rightarrow\infty$ is not fixed. So, from the variable end-point condition, we have
% --------------------------------------------------------------------------------
\begin{equation}
\label{eq:vhz}
\frac{\dd H}{\dd k'_r}\Big|_{r\rightarrow\infty} = \left(\frac{\rho}{p_r} - \frac{\rho}{ p_t} - 2 f_1 r\right)\Big|_{r\rightarrow\infty} = 0.
\end{equation}
From equation~(\ref{eq:steq3}), we can determine $C_1=0$. Thus, the Lagrangian multiplier $f_1$ is
% ---------------------------------------------------------------------------------
\begin{equation}
\label{eq:lmf1}
f_1 = \frac{\rho}{2 r p_r} - \frac{\rho}{2 r p_t}.
\end{equation}
From equations~(\ref{eq:steq2}) and (\ref{eq:steq4}), we can obtain the other Lagrangian multiplier $f_2$ as
% ---------------------------------------------------------------------------------
\begin{equation}
\label{eq:lmf2}
f_2 = \frac{\rho}{2p_t} - \lambda,
\end{equation}
where $\lambda$ is an integration constant. From equations~(\ref{eq:steq2}) and (\ref{eq:lmf1}), we have
% --------------------------------------------------------------------------------
\begin{equation}
\label{eq:2ndese1}
\frac{\dd}{\dd r} (\frac{\rho}{p_t})  = \frac{\rho}{r p_r} - \frac{\rho}{r p_t},
\end{equation}
which can be further expressed as
% --------------------------------------------------------------------------------------
\begin{equation}
\label{eq:2ndese2}
\frac{\dd\ln\rho}{\dd r}-\frac{\dd\ln p_t}{\dd r}-\frac{p_t}{r p_r}+\frac{1}{r}=0.
\end{equation}
If we substitute the two Lagrangian multipliers into equation~(\ref{eq:steq1}), and use the Jeans equation~(\ref{eq:2ndoe}) and equation~(\ref{eq:2ndese2}), we can simplify equation~(\ref{eq:steq1}) as
% --------------------------------------------------------------------------------------
\begin{equation}
\label{eq:1stese}
\frac{3}{2}\frac{\dd \ln\rho}{\dd r} - \frac{1}{p_t} \frac{\dd p_r}{\dd r}-\frac{2 p_r}{r p_t} + \frac{2}{r} = \lambda\frac{G m}{r^2}.
\end{equation}
If the virialization relation, equation~(\ref{eq:vir1}) or (\ref{eq:vir2}), is not included as an extra constraint, then in this case $f_1=0$. So, $f_2$ is just a constant, and from equation~(\ref{eq:steq3}) or equation~(\ref{eq:lmf1}), we have $p_r = p_t$. Hence, equation~(\ref{eq:steq1}) degenerates to
% --------------------------------------------------------------------------------------
\begin{equation}
\label{eq:oldese}
\frac{3}{2}\frac{\dd \ln p_r}{\dd r} - \frac{5}{2}\frac{\dd \ln \rho}{\dd r} = - \lambda \frac{Gm}{r^2},
\end{equation}
which is exactly the equation that we derived in a previous study \citep{hep10}. According to our latest results, this equation should not be accurate. It can work well only in combination with the Jeans equation, and also with an empirical density profile (e.g. the Einasto form).

To summarize, with the truncated DF of equation~(\ref{eq:3df}), the complete set of variables contains only the density profile $\rho(r)$ and the radial and one-dimensional tangential pressure $p_r(r)$ and $p_t(r)$. The Jeans equation~(\ref{eq:2ndoe}) and equations~(\ref{eq:2ndese2}) and (\ref{eq:1stese}), derived from our statistical-mechanical approach, constitute the complete equation system for the three variables $\rho$, $p_r$ and $p_t$. As analysed in Appendix~\ref{sec:appdx}, a separable DF such as equation~(\ref{eq:3df}) will not yield correct results. However, if the theoretical framework of the statistical mechanics that we have formulated is correct, we can expect that the predictions will be more and more accurate compared with the simulation results, as more and more expansion coefficients of $F(\vc{x}, \vc{v})$ in equation~(\ref{eq:lmpd2}) are included.

We see that, in our statistical-mechanical approach, as more and more moments and moment equations are involved, more and more expansion coefficients are needed to express the DF of equation~(\ref{eq:lmpd2}). Thus, we can obtain any possible velocity distribution with the Boltzmann-Gibbs statistics. Hence, the idea that Boltzmann-Gibbs statistics can only produce a canonical distribution, such as a Maxwell-Boltzmann distribution, is not correct \citep[cf.][]{hansen05}. \citet{kafri09} presented a comment on non-extensive statistical mechanics \citep{tsallis88}, and pointed out that the Boltzmann-Gibbs statistics can also yield a long tail (i.e. power law) distribution. Additionally, it might also be unnecessary to invoke the general $H$-function other than the Boltzmann-Gibbs entropy \citep[cf.][]{tremaine86}. See also the discussion by \citet{bouchet06} about the validity of Boltzmann-Gibbs statistics when using large deviation tools.

% --------------------------------------------------------------------------------------
\section{Discussions and conclusions}
\label{sec:concl}

The statistical mechanics of self-gravitating systems is a long-standing puzzle, which has not yet been successfully solved. In this paper, we have formulated a systematic theoretical framework of the statistical mechanics of spherically symmetric self-gravitating systems, using approaches that differ significantly from the conventional statistical mechanics of short-range interaction systems.

The equilibrium states of self-gravitating systems also consist of mechanical equilibria. By multiplying the steady-state CBE with various powers of velocity, and integrating over all velocities, we obtain the moment equations of various orders. In Appendix~\ref{sec:appdx}, we demonstrate that the DF $f$ is inseparable. As a result, the moments are infinitely many and independent of each other. These moments constitute the endless moment equations~(\ref{eq:meq2}), which characterize the great complexity of the mechanical equilibria. The moment equations can be further transformed into the generalized virialization equations~(\ref{eq:exvir}), and hence mechanical equilibria are actually identical to the generalized virial equilibria.

One subtlety that should be pointed out is that the DF $f$, which satisfies the CBE, is the fine-grained DF. All the velocity moments are derived from this fine-grained DF, whereas the important DF for statistical mechanics is the coarse-grained DF. Although the coarse-grained DF does not obey the CBE, fortunately, we find that the coarse-grained moments differ from the fine-grained moments only by up to second-order small quantities determined by the size of the finite volume element. So, we can safely use these moment equations to characterize the mechanical equilibria of the system.

Ergodicity, or the equiprobability principle, is fundamental for equilibrium statistical mechanics. However, ergodicity is invalid for the whole collisionless self-gravitating system, because, for example, a gravitating particle moves in a static gravitational field along a specific curve, on which the energy and angular momentum of the particle are conserved. Such a regular motion of the particle is definitely not an ergodic motion, in that the particle cannot traverse any place in the system, and also cannot attain any value of velocity. Even if the particles are treated as indistinguishable, the validity of global ergodicity cannot be recovered, either. According to the ensemble theory, the probability distribution of the microstates of a closed system at a fixed energy is derived by assuming that all states on the energy hypersurface in phase space have equal probability. However, the hypersurface of a constant energy for an isolated self-gravitating system is unbounded. Thus, the microcanonical probability distribution cannot be defined, which implies that ergodicity, or equiprobability, is broken.

It seems to be a disaster if global ergodicity is invalid, because the equilibrium statistical mechanics cannot exist without the validity of ergodicity. Fortunately, ergodicity can be re-established locally in the following manner. Within an imagined finitely small volume element of the system, none of the collisionless gravitating particles can be tracked, because these particles are not confined within this limited volume element. Hence, these particles should be treated as indistinguishable. At some instant, these particles, randomly and homogeneously populated in the element, exhibit a static distribution pattern. These different patterns of different instants naturally constitute an ensemble. When the system is in its equilibrium states, the background field quantities (i.e. all the velocity moments) are local invariants, which define a hypersurface for this volume element. It is reasonable to assume that every member of the ensemble of the local distribution patterns occurs with equal probability in this hypersurface. Intuitively, if within the finitely small element, the particles are randomly and homogenously distributed, and the velocity distribution is independent of spatial coordinates, then the local ergodicity should be valid and can be re-established.

We assume that the minimum phase-space element (i.e. the microcell) can accommodate only one gravitating particle, as it is impossible for two gravitating particles to be close to each other but simultaneously to be relatively static. Because of the local ergodicity, the particles will populate these microcells with equal probability, which also indicates that all the microstates for this element occur with equal probability. So, the microstates of the indistinguishable gravitating particles within this volume element should be counted with Fermi-Dirac-like statistics. Moreover, because both the direct gravitational interaction and the statistical correlation between any two gravitating particles can be neglected, which is exactly the same reason to justify the CBE, it is reasonable to assert that the microstates of one volume element do not depend on those of any other elements (i.e. local microstates are spatially independent). Hence, the total microstate number of the whole system can be expressed as the product of the microstate number of all the elements. Based on the local ergodicity, and by using Fermi-Dirac-like statistics, with a non-degenerate condition and the spatial independence of the local microstates, we rederive the Boltzmann-Gibbs entropy, which is exactly the correct entropy form suitable for collisionless self-gravitating systems. Finally, the Boltzmann-Gibbs entropy is additive and hence extensive, which means that the non-extensive \citet{tsallis88} entropy as well as the general $H$-function \citep{tremaine86} are not necessary.

The equilibrium states of self-gravitating systems consist of both mechanical/virial and statistical equilibria, of which the latter should be characterized by the statistical equilibrium equations, derived from entropy principle. Apart from the usual constraints of mass and energy conservation, we demonstrate that the series of moment equations~(\ref{eq:meq2}) or their equivalent virialization equations~(\ref{eq:exvir}) are not only employed to characterize the mechanical/virial equilibria, but also act as additional constraints on the system's entropy functional when performing the variational calculus. This approach is an extension of the original prescription by \citet{white87}, who just used the isotropic Jeans equation as an extra constraint.

We further illustrate this statistical mechanical approach with a truncated, separable DF. In this case, the complete set of variables can be chosen to be the density profile and the radial and one-dimensional tangential pressures. The two equations~(\ref{eq:2ndese2}) and (\ref{eq:1stese}), derived from the entropy principle, together with the Jeans equation~(\ref{eq:2ndoe}), constitute the complete equation system for the three variables. As analyzed in Appendix~\ref{sec:appdx}, such a separable DF will not yield correct results. However, if the theoretical framework of the statistical mechanics that we have formulated is correct, we can expect the predictions to be more and more accurate, compared with simulation results, because more and more expansion coefficients of the DF are included. As a result, we can produce any possible velocity distribution using the Boltzmann-Gibbs statistics. Hence, the idea that Boltzmann-Gibbs statistics can only produce a canonical distribution, such as a Maxwell-Boltzmann distribution, is not correct.

We also discuss the problems of negative specific heat and ensemble inequivalence. We propose that these problems might originate from the misuse of ensemble theory. This is because ensemble theory is established only upon the validity of global ergodicity, which is not valid for self-gravitating systems. As a result, the relationships about the heat capacity of equations~(\ref{eq:cv1}) and (\ref{eq:cv2}) are rootless for self-gravitating systems. Without the validity of ensemble theory, these two relationships should be, at most, considered as two different definitions of the `heat capacity' for self-gravitating systems.

We have formulated a systematic theoretical framework of the statistical mechanics of spherically symmetric self-gravitating systems, within which we encounter many difficult mathematical problems, such as the evaluation of both the entropy functional and the velocity moments, and the variational calculus of deriving the equations of statistical equilibria. These will be the tasks of future investigations.

% -------------------------------------------------------------------------------------
\section*{Acknowledgements}

The author would like to thank Dr A. Ludlow and Dr J. Navarro for providing their Aquarius simulation data. The author is also very grateful for a discussion with P. H. Chavanis, Y. Levin and V. Gurzadyan, and for the suggestions and comments made by the anonymous referee. This work is supported by the National Basic Research Programme of China (No:2010CB832805).

% -------------------------------------------------------------------------------------

%-----------------------------------------------------------------------------------
%\onecolumn
\appendix
\section{Collisionless Boltzmann equation, velocity moments and moment equations} \label{sec:appdx}

%------------------------------------------------------------------------------------------
The CBE in spherical coordinates $(r,\theta, \phi)$ is \citep{galdyn08}
% --------------------------------------------------------------------------------------
% collisionless Boltzmann equation for spherical system
\begin{equation}
\label{eq:cbes}
\frac{\partial f}{\partial t} + \dot{r}\frac{\partial f}{\partial r} + \dot{\theta} \frac{\partial f}{\partial \theta} + \dot{\phi}\frac{\partial f}{\partial \phi} + \dot{v}_r\frac{\partial f}{\partial v_r} + \dot{v}_{\theta}\frac{\partial f}{\partial v_{\theta}} + \dot{v}_{\phi}\frac{\partial f}{\partial v_{\phi}} = 0,
\end{equation}
where the time derivatives of the coordinates can be expressed in terms of the velocity components:
%-----------------------------------------------------------------------------------------
\begin{equation}
\label{eq:vel}
\dot{r} = v_r, {\hskip 1cm} \dot{\theta} = \frac{v_{\theta}}{r}, {\hskip 1cm} \dot{\phi} = \frac{v_{\phi}}{r \sin\theta}.
\end{equation}
The Lagrange equations give the components of the acceleration:
%-----------------------------------------------------------------------------------------
\begin{displaymath}
\dot{v}_r = \frac{v^2_{\theta}+v^2_{\phi}}{r} - \frac{\partial\Phi}{\partial r},
\end{displaymath}
\begin{displaymath}
\dot{v}_{\theta} = \frac{v^2_{\phi}\cot\theta - v_r v_{\theta}}{r} - \frac{1}{r} \frac{\partial \Phi}{\partial \theta},
\end{displaymath}
\begin{equation}
\label{eq:acc}
\dot{v}_{\phi} = \frac{-v_{\phi}v_r - v_{\phi} v_{\theta}\cot\theta}{r} - \frac{1}{r\sin\theta}\frac{\partial \Phi}{\partial \phi}.
\end{equation}
For a spherically symmetric system in its steady state, all the terms in equation~(\ref{eq:cbes}), including $\partial/\partial t$, $\partial/\partial \theta$ and $\partial/\partial \phi$, should be vanishing. Furthermore, $\partial\Phi / \partial \theta$ and $\partial\Phi / \partial\phi$ in equation~(\ref{eq:acc}) should also be zero.

With all these to hand, we can multiply the steady-state CBE by various powers of velocity, and integrate over all velocities, to obtain a series of moment equations.

For a spherical system, the fine-grained DF $f(\bf x, v)$ reduces to the form of $f(r, v_r, v_{\theta}, v_{\phi})$, where $v_r$ is the radial velocity and $v_{\theta}$ and $v_{\phi}$ are two tangential velocities. The various orders of velocity moments, such as $\rho(r)$, $\rho\overline{v^n_r}$, $\rho\overline{v^n_{\theta}}$,  $\rho \overline{v^n_{\phi}}$, $\rho\overline{v^n_r v^m_{\theta}}$, $\rho\overline{v^n_r v^m_{\phi}}$ and $\rho\overline{v^n_{\theta} v^m_{\phi}}$, can be uniformly defined as
% -------------------------------------------------------------------------------------
\begin{equation}
\label{eq:vm}
\rho\overline{v^k_r v^m_{\theta} v^n_{\phi}} \equiv \int v^k_r v^m_{\theta} v^n_{\phi} f(r, v_r, v_{\theta}, v_{\phi}) \dd v_r \dd v_{\theta} \dd v_{\phi}.
\end{equation}
Here, $k,m,n$ are non-negative even integers (i.e. $k, m, n = 0, 2, 4, \ldots$).

If we multiply the steady-state CBE by $v^{k-1}_r v^m_{\theta} v^n_{\phi}$ and integrate over all velocities, we obtain
% -------------------------------------------------------------------------------------
\begin{displaymath}
\frac{\dd}{\dd r}(\rho\overline{v^k_r v^m_{\theta} v^n_{\phi}}) + \frac{(m+n+2)}{r} \rho\overline{v^k_r v^m_{\theta} v^n_{\phi}}
\end{displaymath}
\begin{displaymath}
- \frac{(k-1)}{r} \rho\overline{v^{k-2}_r v^{m+2}_{\theta} v^n_{\phi}} - \frac{(k-1)}{r} \rho\overline{v^{k-2}_r v^m_{\theta} v^{n+2}_{\phi}}
\end{displaymath}
\begin{displaymath}
- \frac{m\cot\theta}{r} \rho\overline{v^{k-1}_r v^{m-1}_{\theta} v^{n+2}_{\phi}} + \frac{(n+1)\cot\theta}{r} \rho\overline{v^{k-1}_r v^{m+1}_{\theta} v^n_{\phi}}
\end{displaymath}
\begin{equation}
\label{eq:home0}
= - (k-1)\frac{\dd \Phi}{\dd r} \rho\overline{v^{k-2}_r v^m_{\theta} v^n_{\phi}},
\end{equation}
which can be further divided into two relations. The first relation is
%--------------------------------------------------------------------------------------
\begin{displaymath}
\frac{\dd}{\dd r}(\rho\overline{v^k_r v^m_{\theta} v^n_{\phi}}) + \frac{(m+n+2)}{r} \rho\overline{v^k_r v^m_{\theta} v^n_{\phi}}
\end{displaymath}
\begin{displaymath}
- \frac{(k-1)}{r} \rho\overline{v^{k-2}_r v^{m+2}_{\theta} v^n_{\phi}} - \frac{(k-1)}{r} \rho\overline{v^{k-2}_r v^m_{\theta} v^{n+2}_{\phi}}
\end{displaymath}
\begin{equation}
\label{eq:meq1}
= - (k-1)\frac{\dd \Phi}{\dd r} \rho\overline{v^{k-2}_r v^m_{\theta} v^n_{\phi}},
\end{equation}
where $k$ should be an even number. In this case, $n$ and $m$ must also be even numbers, with $k \geq 2$, and $m, n \geq 0$. The second relation is the following recursive relation
%----------------------------------------------------------------------------------------
\begin{equation}
\label{eq:receq}
m \overline{v^{k-1}_r v^{m-1}_{\theta} v^{n+2}_{\phi}} = (n+1)\overline{v^{k-1}_r v^{m+1}_{\theta} v^n_{\phi}},
\end{equation}
where $k, m$ are odd numbers, with $k, m \geq 1$, and $n$ is an even number, with $n \geq 0$. Using equation~(\ref{eq:receq}) repeatedly, we have
% ------------------------------------------------------------------------------------
\begin{displaymath}
\overline{v^{k-1}_r v^{m-1}_{\theta} v^{n+2}_{\phi}} = \frac{(n+1)}{m} \overline{v^{k-1}_r v^{m+1}_{\theta} v^n_{\phi}}
\end{displaymath}
\begin{displaymath}
= \frac{(n+1)(n-1)}{m(m+2)} \overline{v^{k-1}_r v^{m+3}_{\theta} v^{n-2}_{\phi}} = ...
\end{displaymath}
\begin{equation}
\label{eq:re1}
= \frac{(m-2)!! (n+1)!!}{(m+n)!!} \overline{v^{k-1}_r v^{m+n+1}_{\theta}},
\end{equation}
or
% ------------------------------------------------------------------------------------
\begin{displaymath}
\overline{v^{k-1}_r v^{m+1}_{\theta} v^n_{\phi}} = \frac{m}{(n+1)} \overline{v^{k-1}_r v^{m-1}_{\theta} v^{n+2}_{\phi}}
\end{displaymath}
\begin{displaymath}
= \frac{m (m-2)}{(n+1)(n+3)} \overline{v^{k-1}_r v^{m-3}_{\theta} v^{n+4}_{\phi}} = ...
\end{displaymath}
\begin{equation}
\label{eq:re2}
= \frac{m!! (n-1)!!}{(m+n)!!} \overline{v^{k-1}_r v^{m+n+1}_{\phi}}.
\end{equation}
%-------------------------------------------------------------------------------------
Equations~(\ref{eq:re1}) and (\ref{eq:re2}) indicate that $\overline{v^{k-1}_r v^{m-1}_ {\theta} v^{n+2}_{\phi}}$ or $\overline{v^{k-1}_r v^{m+1}_{\theta} v^n_{\phi}}$ are not independent. From equations~(\ref{eq:receq}), (\ref{eq:re1}) and (\ref{eq:re2}), with the replacement $k-1 \rightarrow k $, and $m+n+1 \rightarrow m$, we can obtain
% -----------------------------------------------------------------------------------
\begin{equation}
\label{eq:eq2}
\overline{v_r^{k} v^{m}_{\theta}} = \overline{v^{k}_r v^{m}_{\phi}},
\end{equation}
where $k, m$ are even numbers ($k, m=0, 2, 4, \ldots $). With equations~(\ref{eq:receq}) -- (\ref{eq:eq2}), and with the replacement $k \rightarrow k+2 $, and $m+n \rightarrow m$, equation~(\ref{eq:meq1}) can be reduced to
% -------------------------------------------------------------------------------------
\begin{displaymath}
\frac{\dd}{\dd r}(\rho\overline{v^{k+2}_r v^m_{t}}) + \frac{(m+2)}{r} \rho\overline{v^{k+2}_r v^m_{t}}
\end{displaymath}
\begin{equation}
\label{eq:meq2}
- \frac{(k+1)}{r}\frac{(m+2)}{(m+1)} \rho\overline{v^k_r v^{m+2}_{t}} = - (k+1)\frac{\dd \Phi}{\dd r} \rho\overline{v^k_r v^m_{t}},
\end{equation}
where $v_t$ is either $v_{\theta}$ or $v_{\phi}$, with $k,m=0,2,4, ...$. Generally, there are $N$ independent $2N$th-order moment equations, with different combinations of $k$ and $m$ in equation~(\ref{eq:meq2}). See also \citet{dejonghe92} and \citet{an11} for the derivation of this equation.

We would like to investigate whether these moment equations can be reduced to their counterparts of lower orders. If the DF is uncorrelative, i.e. $f$ can be separated as
% --------------------------------------------------------------------------------------
\begin{equation}
\label{eq:fsep}
f(v_r, v_{\theta}, v_{\phi}) = f_1(v_r) f_2(v_{\theta}) f_3(v_{\phi}),
\end{equation}
where $f_1$, $f_2$ and $f_3$ are one-dimensional DFs, then the velocity moments of equation~(\ref{eq:vm}) can also be separated as
\begin{displaymath}
\overline{v^k_r v^m_{\theta} v^n_{\phi}\vphantom{v^k}} = \overline{v^k_r\vphantom{v^k}}\ \overline{v^m_{\theta} \vphantom{v^k}}\ \overline{v^n_{\phi}\vphantom{v^k}}.
\end{displaymath}
Setting $n=0$, and with the replacement $m \rightarrow 2m+1$, from equation~(\ref{eq:receq}) we obtain
% --------------------------------------------------------------------------------------
\begin{eqnarray}
\label{eq:uncoeq1}
\overline{v^{2m+2}_\theta}
& = & (2m+1)\overline{v^{2m}_\theta} \ \overline{v^2_\phi} \nonumber \\
& = & (2m+1)(2m-1)\overline{v^{2m-2}_\theta} \ (\overline{v^2_\phi})^2 \nonumber \\
&   & \vdots \nonumber \\
& = & (2m+1)!! (\overline{v^2_\phi})^{m+1} \nonumber \\
& = & (2m+1)!! (\overline{v^2_\theta})^{m+1}.
\end{eqnarray}
Generally, with $v_t$ being either $v_{\theta}$ or $v_{\phi}$, we have
\begin{equation}
\label{eq:uncoeq2}
\overline{v^{2m+2}_t}= (2m+1)!! (\overline{v^2_t})^{m+1},
\end{equation}
which means that the moments of $v_t$ are not independent. In this case, using the Jeans equation~(\ref{eq:2ndoe}), the 4th-order moment equation~(\ref{eq:4thoe2}) can be reduced to
% ------------------------------------------------------------------------------------
\begin{equation}
\label{eq:rd4th1}
\overline{v^2_r}\frac{\dd}{\dd r}(\overline{v^2_{t}}) + \frac{2}{r} \overline{v^2_{t}} (\overline{v^2_r} - \overline{v^2_{t}}) = 0.
\end{equation}
This can be re-expressed as
% ------------------------------------------------------------------------------------
\begin{equation}
\label{eq:rd4th2}
\frac{\dd\ln \overline{v^2_{t}}}{\dd \ln r} = -2 \left( 1 - \frac{\ \overline{v^2_t}\ }{\ \overline{v^2_r}\ } \right) = -2 \beta,
\end{equation}
where $\beta$ is the anisotropy parameter. From the simulations by \citet{navarro10} and \citet{ludlow10}, we know that $\beta$ increases with the radius $r$ from zero to the maximum, and then decreases to about zero again at the outskirts of the dark halos (i.e. $\beta \geq 0$ for most of the radii, except for the possible negative value at the outer boundary). Hence, $\overline{v^2_t}$ resulting from equation~(\ref{eq:rd4th2}) should monotonically decrease with $r$ for most of the radii, which is in contrast to the same simulation results of \citet{navarro10} and \citet{ludlow10}. This suggests that equation~(\ref{eq:rd4th2}) and equations~(\ref{eq:uncoeq1}) -- (\ref{eq:rd4th1}) are not correct. Hence, the DF $f$ is inseparable, which implies that these moment equations cannot be reduced to the lower-order counterparts.

% --------------------------------------------------------------------------------------
\label{lastpage}
\end{document}